\documentclass[notitlepage,a4]{article}
\usepackage{amsmath}
\usepackage{amssymb}
\usepackage{labelfig}
\usepackage{graphicx}
\usepackage{titling}

\newcommand{\taub}{\mbox{\boldmath{$\tau$}}}
\newcommand{\sigmab}{\mbox{\boldmath{$\sigma$}}}
\newcommand{\xib}{\mbox{\boldmath{$\xi$}}}

\newcommand{\szero}{^{(0)}}

\newcommand{\bquote}{\begin{quote}\footnotesize}
\newcommand{\equote}{\end{quote}}

\def \d2xdt2{{\frac{d^{2}x}{dt^{2}}}}
\def \d2ydt2{{\frac{d^{2}y}{dt^{2}}}}

\newcommand{\btwo}{\left[ \begin{array}{cc}}
\newcommand{\etwo}{\end{array} \right]}
\newcommand{\bthree}{\left[ \begin{array}{ccc}}
\newcommand{\ethree}{\end{array} \right]}
\newcommand{\bfour}{\left[ \begin{array}{cccc}}
\newcommand{\efour}{\end{array} \right]}
\newcommand{\bvec}{\left[ \begin{array}{c}}
\newcommand{\evec}{\end{array} \right]}

\newcommand{\half}{{1/2}}

\newcommand{\e}{\begin{equation}}
\newcommand{\ee}{\end{equation}}

\newcommand{\eps}{{\varepsilon}}
\newcommand{\Oh}{{\mathcal{O}}}

\newcommand{\junk}[1]{}
\newcommand{\hide}[1]{{}}  

\renewcommand{\e}{\begin{equation}}
\renewcommand{\ee}{\end{equation}}
\newcommand{\xb}{{\mathbf{x}}}

\newcommand{\R}{\mathbb{R}}

\newcommand{\options}{\left\{  \begin{array}{ccc}}
\newcommand{\finishoptions}{ \end{array} \right.}

\newcommand{\Db}{\mathbf{D}}

\newcommand{\E}{\mathcal{E}}

\newcommand{\registered}{{\ooalign{\hfil\raise .00ex\hbox{\scriptsize R}\hfil\crcr\mathhexbox20D}}}

\def\u#1{\mbox{\boldmath $#1$}}
\def\d#1{\partial_{#1}}
\def\eqref#1{(\ref{#1})}

\def\szero{^{(0)}}

\newcommand{\fracds}{\displaystyle\frac}
\newcommand{\proofend}{\vrule height 6pt width 6pt depth 1pt}

\newcommand{\bfive}{\left[ \begin{array}{ccccc}}
\newcommand{\efive}{\end{array}\right]}

\everymath{\displaystyle}

\pretitle{\begin{center}\huge\bfseries}
\posttitle{\par\end{center}\vskip 0.5em}
\preauthor{\begin{center}\Large}
\postauthor{\end{center}}
\predate{\begin{center}\noindent $^{1}$School of Mathematics and Manchester Centre for Nonlinear Dynamics, University of Manchester, Manchester, UK\\$^2$Mathematics Department, Duke University, Durham, NC, USA\\$^3$Department of Mathematics, North Carolina State University, Raleigh, NC, USA\end{center}\par}
\postdate{}

\title{Well-posed continuum equations for granular flow with compressibility and $\mu(I)$-rheology}
\author{T. Barker$^1$, D.G. Schaeffer$^2$, M. Shearer$^3$ \& J.M.N.T Gray$^1$}
\date{}
\begin{document}

\maketitle
\thispagestyle{empty}

\begin{abstract}\normalsize
\noindent Continuum modelling of granular flow has been plagued with the issue of ill-posed equations for a long time. Equations for incompressible, two-dimensional flow based on the Coulomb friction law are ill-posed regardless of the deformation, whereas the rate-dependent $\mu(I)$-rheology is ill-posed when the non-dimensional strain-rate $I$ is too high or too low. Here, incorporating ideas from Critical-State Soil Mechanics, we derive conditions for well-posedness of PDEs that combine compressibility with $I$-dependent rheology. When the $I$-dependence comes from a specific friction coefficient $\mu(I)$, our results show that, with compressibility, the equations are well-posed for all deformation rates provided that $\mu(I)$ satisfies certain minimal, physically natural, inequalities.   
\end{abstract}

\section{Introduction}

Much effort has been devoted to formulating constitutive laws for continuum models of granular materials \cite{Jackson1983, Schaeffer1987, SchaefferPitman1988, GDRMidi2004, JopForterrePouliquen2006}. However, the lack of acceptable dynamic theories, i.e., well posed equations  in the sense of Joseph \& Saut \cite{JosephSaut1990}, for granular flow has severely hampered progress in modelling many geophysical and industrial problems. In the simplest class of models, flow is described by Partial Differential Equations (PDEs) for the density, the velocity vector and the stress tensor; conceptually, such models are hardly more complicated than the Navier--Stokes equations.  The equations represent conservation laws for mass and momentum coupled to constitutive equations to close the system. However, despite the appeal of their simplicity, they have been plagued with ill-posedness, i.e. small perturbations grow at an unbounded rate in the limit that their wavelength tends to zero \cite{JosephSaut1990}. Such behaviour is clearly unphysical. However, the immediate practical implication of ill-posedness is that numerical computations either blow-up, even at finite resolution, or do not converge to a well-defined solution as the grid is refined, i.e. the numerical results are grid dependent \cite{Gray1999, Woodhousetal2012, Barker2015}.

The first model of this type \cite{Schaeffer1987,Mandel1964, PitmanSchaeffer1987} specifies constitutive laws that 
represent a tensorial generalisation of the work of de Coulomb \cite{Coulomb1773} on earthwork fortifications. In 
the language of plasticity theory, it is a rate-independent, rigid/perfectly-plastic model with a yield condition 
based on friction between the grains. However, it was shown to be ill-posed in all two-dimensional contexts and 
all realistic three-dimensional contexts \cite{Schaeffer1987}. Critical State Soil Mechanics \cite{Jackson1983} is 
a sophisticated elaboration of Coulomb behaviour that allows for compressibility. It also suffers from 
ill-posedness, depending of the degree of consolidation. This ill-posedness is much less severe than for a Coulomb 
material \cite{PitmanSchaeffer1987, SchaefferPitman1988}, but still bad enough to block its use in applications. 
More recently, the $\mu(I)$-rheology \cite{GDRMidi2004, DaCruzEtAl2005, JopForterrePouliquen2006} introduces a 
modest amount of rate dependence into (incompressible) Coulomb behaviour through the non-dimensional 
\emph{inertial number}, which is proportional to the shear-rate and inversely proportional to the square-root of 
the pressure. As shown in Barker \textit{et al.} \cite{Barker2015}, this theory leads to well posed 
(two-dimensional) equations in a significant region of state space, but it is ill-posed at both low and high 
inertial numbers.

This paper is centred on formulating constitutive equations that extend the incompressible $\mu(I)$-rheology of Jop 
\textit{et al.} \cite{JopForterrePouliquen2006} to compressible deformations, by combining it with Critical State 
Soil Mechanics. The main result is that in two dimensions, the new model is well-posed for all densities, for all 
stress states, and for all deformation rates. In other words, to obtain well-posedness, we modify Coulomb behaviour 
by including only two natural, fairly small, perturbations of the theory, namely compressibility and rate 
dependence. This has the advantage that it retains the conceptual simplicity of the original theory. Although we 
consider only two-dimensional flow, it should be noted that in numerous cases it has been found that flow in two 
dimensions is more prone to ill-posedness than in three \cite{Schaeffer1987,SchaefferPitman1988, Pitman1988}. 
Thus, we anticipate that the corresponding three-dimensional equations including these effects will also be well 
posed.

Currently a wide range of new constitutive laws for granular materials are being developed including the $\mu(I)$-rheology \cite{GDRMidi2004, JopForterrePouliquen2006}, elasto-plastic formulations \cite{Kamrin2010, JiangLiu2007}  non-local rheologies \cite{PouliquenForterre2009, Kamrin2012, Kamrin2015, Bouzid2013}, kinetic theory \cite{JenkinsSavage1983}, as well as Cosserat \cite{Harris2005}, micro-structural \cite{SunSundaresan2011}  and hypoplastic theories \cite{Wu1996}. Enormous progress has been made over the past decade and there is the realistic and exciting prospect that practical granular flows, that span the solid-like, liquid-like and gaseous regimes, may shortly be described by continuum models. In this paper we seek to understand one of the conceptually simplest formulations that leads to well-posed equations.

In Section~2 we introduce the equations to be studied and formulate our well-posedness result for them. This 
theorem is proved in Sections~3 and 4.  In two appendices we summarise key ideas from Critical State Soil Mechanics 
and survey topics regarding ill-posed partial differential equations.

\section{Governing equations}

Dense granular flow is described by the solids-volume fraction $\phi$, the velocity vector $\u{u}$, and the stress tensor $\u{\sigma}$. In two dimensions this constitutes six scalar unknowns that are spatially and temporally dependent. These are governed by conservation laws plus constitutive relations. Conservation of mass gives the scalar equation
\begin{equation} \label{c-mass}
  (\partial_t +u_j \partial_j ) \phi +\phi \mbox{ div } \u{u} =0 \, ,
\end{equation}
and conservation of momentum gives the vector equation
\begin{equation} \label{c-mom}
  \rho_*\phi(\partial_t+u_j \partial_j) u_i= \partial_j \sigma_{ij}+\rho_*\phi g_i \, ,
\end{equation}
where $\rho_*$ is the constant intrinsic density and $\mathbf{g}$ is the acceleration due to gravity.
Closure of these equations is achieved through three constitutive relations. 

\subsection{The Coulomb constitutive model}
\begin{figure}
\begin{center}
\SetLabels 
       \endSetLabels
       \strut\AffixLabels{ }
\SetLabels 
	 \L (0.0*0.9) $(a)$\\
	\L (0.5*0.9) $(b)$\\
      \L (0.05*0.748) Minor \\
	\L (0.03*0.7) stress axis\\
	      \L (0.39*0.748) Major \\
	\L (0.37*0.7) stress axis\\
      \L (0.55*0.758) Expansive \\
	\L (0.55*0.71) strain-rate\\
	      \L (0.88*0.758) Contractive \\
	\L (0.89*0.71) strain-rate\\
       \endSetLabels
       \strut\AffixLabels{
	\includegraphics[width=0.45\textwidth]{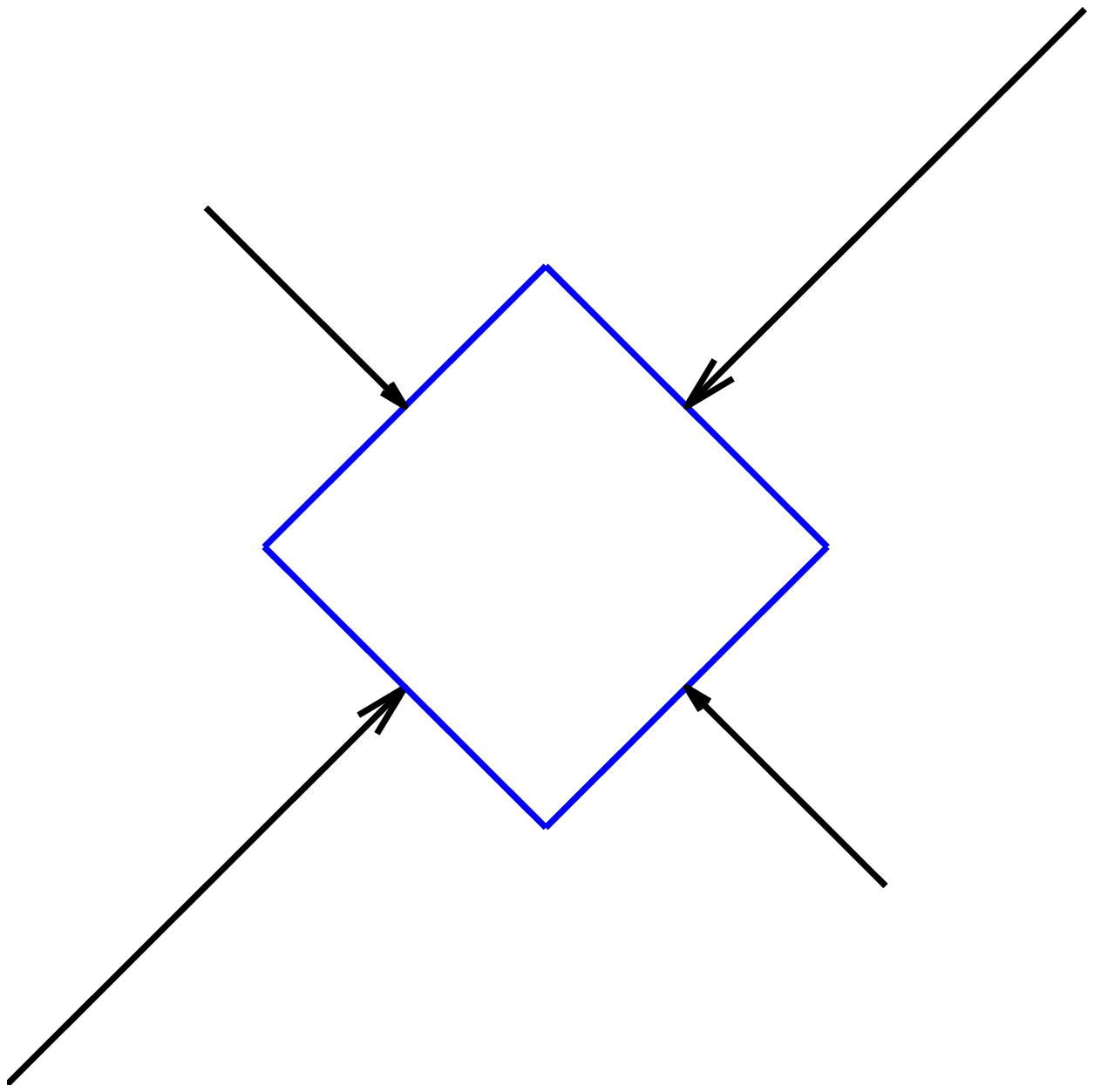}
	\includegraphics[width=0.45\textwidth]{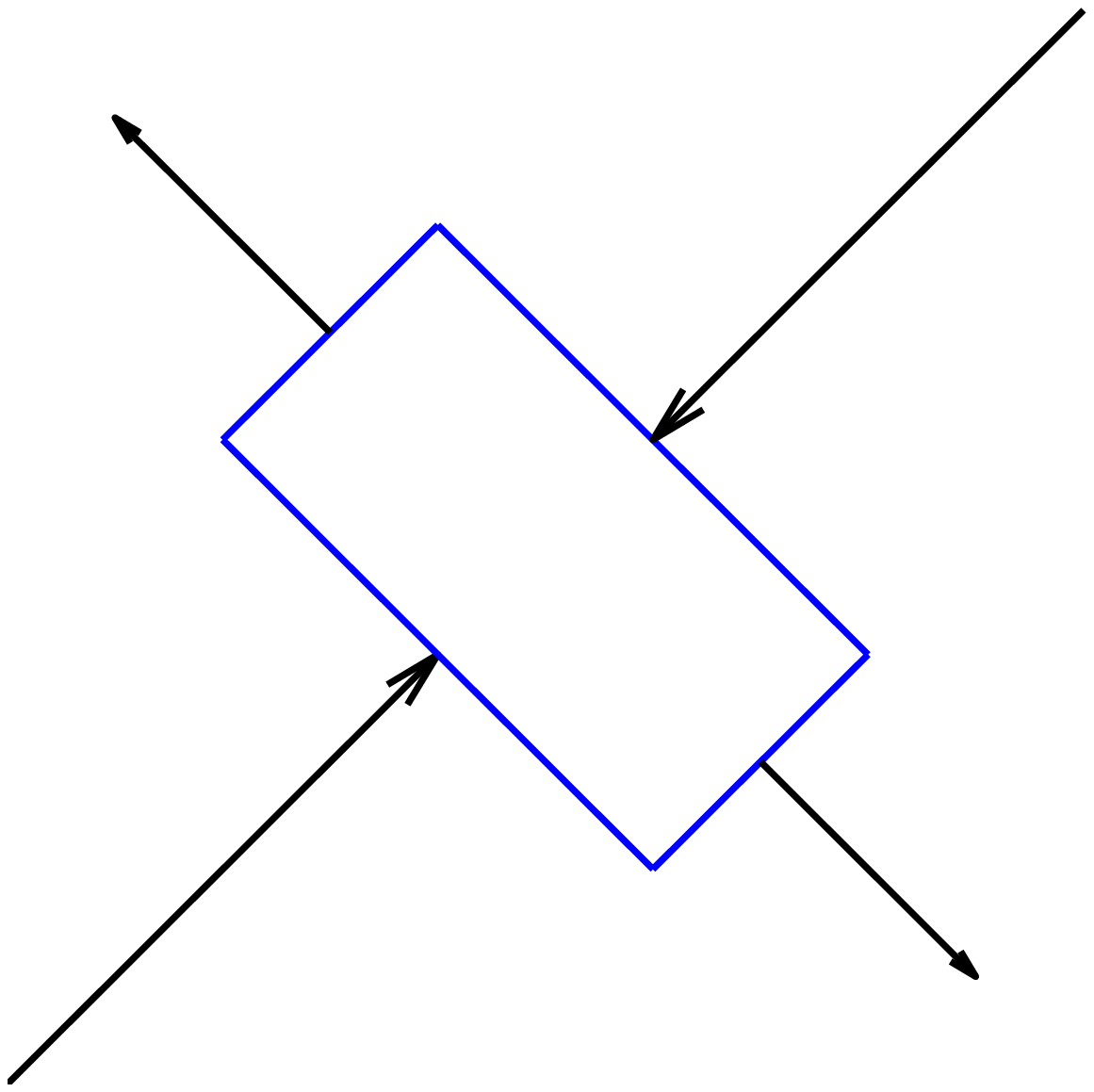}
}
\caption{ \label{fig01} $(a)$ Illustrative stress eigenvectors; along the major axis the stress eigenvalue is $-(p+\|\u{\tau}\|)$ with the minus sign indicating compression. $(b)$ A possible material deformation that is consistent with the stress field in $(a)$.}
\end{center}
\end{figure}

For a Coulomb material, which is assumed to be incompressible, the first constitutive relation states that $\phi$ is a constant. This then reduces \eqref{c-mass} to the
\begin{equation} \label{incompress}
 \mbox{\bf Flow rule:} \qquad \mbox{div } \u{u} =0 \, .
\end{equation}
For the next constitutive relation the stress tensor
\begin{equation} \label{dfn-tau}
 \sigma_{ij} = -p \delta_{ij} +\tau_{ij} \, ,
\end{equation}
is decomposed into a pressure term (where $p=-\sigma_{ii}/2$) plus a trace-free tensor $\u{\tau}$, called the deviatoric stress. The second relation is then the
\begin{equation} \label{yield-cond-coulomb}
 \mbox{\bf Yield condition:}  \qquad  \| \u{\tau} \| = \mu p \, ,
\end{equation}
where $\mu$ is a constant and for any tensor $\u{T}$ the norm is defined by
\begin{equation} \label{dfn-norm}
 \| \u{T} \| = \sqrt{T_{ij}T_{ij}/2} \, .
\end{equation}
This yield condition expresses the idea that a granular material cannot deform unless the shear stress is
sufficient to overcome friction\footnote{Thus, \eqref{yield-cond-coulomb} contains the implicit  assumption that
material is actually deforming.  Otherwise \eqref{yield-cond-coulomb} must be replaced by inequality, $ \| \u{\tau}
\| \le \mu p$, and the governing equations are underdetermined unless further relations, such as those of 
elasticity, 
are included.}. The third constitutive relation requires that the
eigenvectors of the deviatoric stress tensor and the deviatoric strain-rate tensor%
\footnote{Note that for
incompressible flow, the
full strain-rate tensor $( \partial_j u_i+\partial_i u_j)/2$ and the deviatoric strain-rate tensor are equal
since the second term on the right in \eqref{dfn-strain-rate} vanishes.}
\begin{equation} \label{dfn-strain-rate}
 D_{ij} = \frac{1}{2} ( \partial_j u_i+\partial_i u_j) - \frac{1}{2}(\mbox{div } \u{u}) \delta_{ij} \, ,
\end{equation}
are aligned (see Figure \ref{fig01} for motivation), which may be written
\begin{equation} \label{align-coulomb}
 \mbox{\bf Alignment:} \qquad \frac{D_{ij}}{\| \u{D} \|} =  \frac{\tau_{ij}}{\| \u{\tau} \|} \, .
\end{equation}
In words \eqref{align-coulomb} may be interpreted as asserting that in the space of trace-free symmetric
$2\times2$ matrices, \emph{which is two-dimensional}, $\u{D}$ and $\u{\tau}$ are parallel. Thus, this matrix
equation entails only one scalar relation.
For reference below we record that
\begin{equation} \label{dev-strain-rate}
 \u{D} = \frac{1}{2} \btwo \partial_1 u_1 -\partial_2 u_2& \partial_1 u_2 + \partial_2 u_1 \\
  \partial_1 u_2 + \partial_2 u_1& \partial_2 u_2 -\partial_1 u_1 \etwo \, .
\end{equation}

It is customary \cite{Schaeffer1987}, \cite{JopForterrePouliquen2006} to process these equations by expressing the deviatoric stress  $\u{\tau}$ in terms of $p$ and the strain rate as follows:
\begin{equation} \label{solve-for-tau}
  \tau_{ij} = \| \u{\tau} \| \frac{\tau_{ij}}{\| \u{\tau} \|} = \mu p \frac{D_{ij}}{\| \u{D} \|} \, ,
\end{equation}
where we have invoked \eqref{yield-cond-coulomb} and \eqref{align-coulomb}. We may substitute \eqref{solve-for-tau} into \eqref{c-mom} to obtain
\begin{equation} \label{coulomb-system-mom}
  \rho_*\phi(\partial_t  +u_j \partial_j) u_i=
 \partial_j\left[\frac{\mu p}{\|\u{D}\|} D_{ij} \right] -\partial_i p +\rho_*\phi g_i \, ,
\end{equation}
and the resulting equation, together with \eqref{incompress}, gives three equations
for pressure $p$ and velocity $\u{u}$.
In form, at least, these equations resemble the incompressible Navier-Stokes equation.
However, in two dimensions (as considered here) they are always ill-posed \cite{Schaeffer1987}.

\subsection{Incompressible $\mu(I)$-rheology}

Work described by the Groupement De Recherche Milieux Divis\'es \cite{GDRMidi2004} has significantly improved the Coulomb model by
including some rate dependence (in the sense of
plasticity \cite{Perzyna1966}) in the yield condition while making no changes in the incompressible flow rule
\eqref{incompress}
and the alignment condition \eqref{align-coulomb}.
Specifically, a wide range of experiments is captured by replacing
the constant $\mu$ in \eqref{yield-cond-coulomb} by an
increasing function $\mu(I)$ of the \emph{inertial number},
\begin{equation} \label{dfn-I}
 I = \frac{2d \| \Db \|}{\sqrt{p/ \rho_*}} \, ,
\end{equation}
where $d$ is the particle diameter.
The expression
\begin{equation} \label{canonical-mu}
 \mu(I) = \mu_1 + \frac{\mu_2-\mu_1}{I_0/I+1} \, ,
\end{equation}
where $\mu_1$, $\mu_2 $ and $I_0$ are constants with $\mu_2>\mu_1$, is a frequently used
form \cite{JopForterrePouliquen2005}.
Below we shall assume that
\begin{equation} \label{cond-on-mu}
 \mu'(I)>0 \qquad \mbox{and} \qquad \mu''(I)<0 \, .
\end{equation}

The modified yield condition changes \eqref{coulomb-system-mom} to read
\begin{equation} \label{mu(I)-system-mom}
  \rho_*\phi(\partial_t  +u_j \partial_j) u_i=
 \partial_j\left[\frac{\mu(I) p}{\|\u{D}\|} D_{ij} \right] -\partial_i p +\rho_*\phi g_i  \, .
\end{equation}
The effect of this seemingly small perturbation is profound. Unlike for Coulomb
material, equations \eqref{mu(I)-system-mom}
and \eqref{incompress} are well-posed for a
significant range of inertial numbers, specifically when the deformation rate is neither too small nor too large
relative to the pressure
\cite{Barker2015}.

\subsection{Compressibility and $I$-dependent rheology}
We refer to Critical State Soil Mechanics (cf. Appendix 1) for guidance in introducing compressibility
into the rheology.
Thus, we make no change in the alignment condition \eqref{align-coulomb};
we assume $\phi$-dependence in the yield condition,
\begin{equation} \label{yield-cond-Cmu}
   \|\taub\| = Y(p,\phi,I) \, ;
\end{equation}
and we allow for volumetric changes by
introducing a new function $f(p,\phi,I)$ and modifying the flow rule to
\begin{equation} \label{flow-rule-Cmu}
\mbox{div }\u{u} =2 f(p,\phi,I) \, \|\Db\| \, .
\end{equation}
To get well posed equations, we require that the yield condition and the flow-rule functions 
are related by the equation%
\footnote{If $Y$ and $f$ are independent of $I$, then \eqref{Cmu-well-posed-cond} leads to the CSSM flow rule
\eqref{flow-rule-CSSM} derived from normality.}
\begin{equation} \label{Cmu-well-posed-cond}
 \frac{\partial Y}{\partial p} - \frac{I}{2p} \frac{\partial Y}{\partial I} =
    f + I \frac{\partial f}{\partial I} \, ,
\end{equation} 
and that they satisfy the 
inequalities
\begin{equation} \label{ineq-for-thm}
 \mbox{(a) }\partial_I Y>0 \qquad \mbox{and}\qquad  \mbox{(b) } \partial_p f -\frac{I}{2p}\partial_I f <0 \, .
\end{equation}

We may now state our main result, the well posedness theorem for the system
\eqref{c-mass}, \eqref{c-mom}, \eqref{align-coulomb}, \eqref{yield-cond-Cmu}, \eqref{flow-rule-Cmu}, which we call
the CIDR equations.  (Mnemonic: compressible $I$-dependent rheology.)
The term \emph{linearly well posed} is defined in Appendix~2,
and the result is proved in Sections~3 and 4.

\bigskip

\noindent{\bf Theorem} \emph{Under hypotheses \eqref{Cmu-well-posed-cond} and \eqref{ineq-for-thm},
 the CIDR system is linearly well posed.}

\bigskip

\emph{Remark:} The  $I$-dependence in these equations need not relate to a friction coefficient $\mu(I)$.
In \S\ref{mui_sec} we connect the equations to $\mu(I)$-rheology.

\subsection{Derivation of evolution equations}

To place the equations in a larger continuum-mechanics context, we show that
the CIDR equations of motion can be rewritten as a system of three evolution equations for the
velocity $\u{u}$ and the solids fraction $\phi$.
In form, these equations are analogous to the Navier-Stokes equations for a viscous,
compressible fluid.
We make no use of this form of the equations in our proof of well-posedness.

We want to eliminate stresses from the equations of motion.
To this end, we propose to solve for the mean stress $p$ using the flow rule \eqref{flow-rule-Cmu}, which
we rewrite as
\begin{equation} \label{flow-rule-Cmu-rewritten}
   f(p,\phi,I)  = \frac{\mbox{div }\u{u}}{2  \|\Db\| } \, .
\end{equation}
Note that $f(p,\phi,I)$ depends on $p$
both directly in its first argument and indirectly through $I=2d\|\u{D}\|/\sqrt{p/\rho_*}$ in its third
argument.
However, 
\begin{equation} \label{partial-der-for-IFT}
 \frac{\partial}{\partial p} [f(p,\phi,2d\|\u{D}\|/\sqrt{p/\rho_*})]
    =\partial_p f -\frac{I}{2p}\partial_I f  \, ,
\end{equation}
which by assumption (\ref{ineq-for-thm}b) is nonzero.
Thus, we may apply the Implicit Function Theorem to \eqref{flow-rule-Cmu-rewritten} to solve
$
  p=P(\nabla\u{u},\phi).\footnote{Note that $P$ in fact depends only on $\mbox{div }\u{u}, \|\u{D}\|$ and $\phi.$}
$
Given this, we may define
$$
T(\nabla\u{u} ,\phi)=Y(P(\nabla\u{u},\phi),\phi,I(\nabla\u{u},\phi))
\qquad \mbox{where} \qquad
I(\nabla\u{u},\phi)=\frac{2d\|\u{D}\|}{\sqrt{P(\nabla\u{u},\phi)/\rho_*}} \, ,
$$
and substitute into conservation of momentum to obtain an equation
\begin{equation} \label{cons-mom-no-expl-stress}
 \rho_*\phi(\partial_t  +u_j \partial_j) u_i=
 \partial_j\left[\frac{T(\nabla\u{u} ,\phi) }{\|\u{D}\|} D_{ij} \right]
-\partial_i [P(\nabla\u{u},\phi) ] +\rho_*\phi g_i \, .
\end{equation}
This equation, along with \eqref{c-mass}, gives a system of three evolution equations for the
velocity $\u{u}$ and the solids fraction $\phi$.

\subsection{Connection to $\mu(I)$-rheology}\label{mui_sec}

\begin{figure}
\begin{center}
\SetLabels 
       \endSetLabels
       \strut\AffixLabels{ }
\SetLabels 
	 \L (0.0*0.9) $(a)$\\
	\L (0.42*0.9) $(b)$\\
      \L (0.02*0.5) $C$\\
      \L (0.2*0.0) $\phi$\\
	      \L (0.25*0.05) $\phi_0$\\
	      \L (0.28*0.05) $\phi_0+\eps$\\
		\L (0.245*0.17) $\phi_1$\\
		\L (0.25*0.39) $\phi_2$\\
		\L (0.255*0.7) $\phi_3$\\
	\L (0.435*0.5) $Y$\\
	\L (0.7*0.0) $p$\\
		\L (0.537*0.445) $B$\\
		\L (0.63*0.4) $A$\\
		\L (0.515*0.16) $\phi_1$\\
		\L (0.665*0.29) $\phi_2$\\
		\L (0.87*0.5) $\phi_3$\\
       \endSetLabels
       \strut\AffixLabels{
	\includegraphics[height=0.35\textwidth]{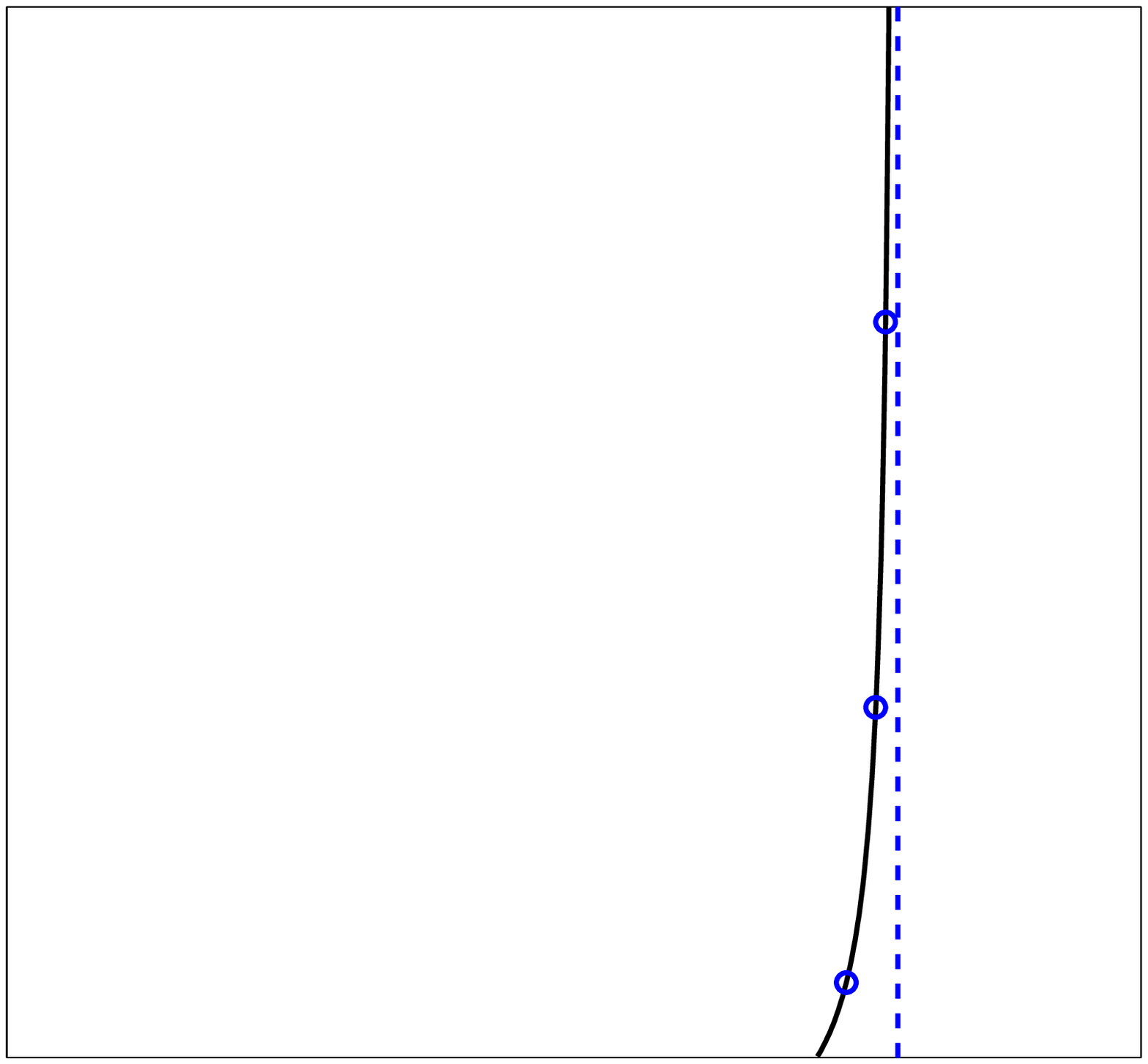}
	\includegraphics[height=0.35\textwidth]{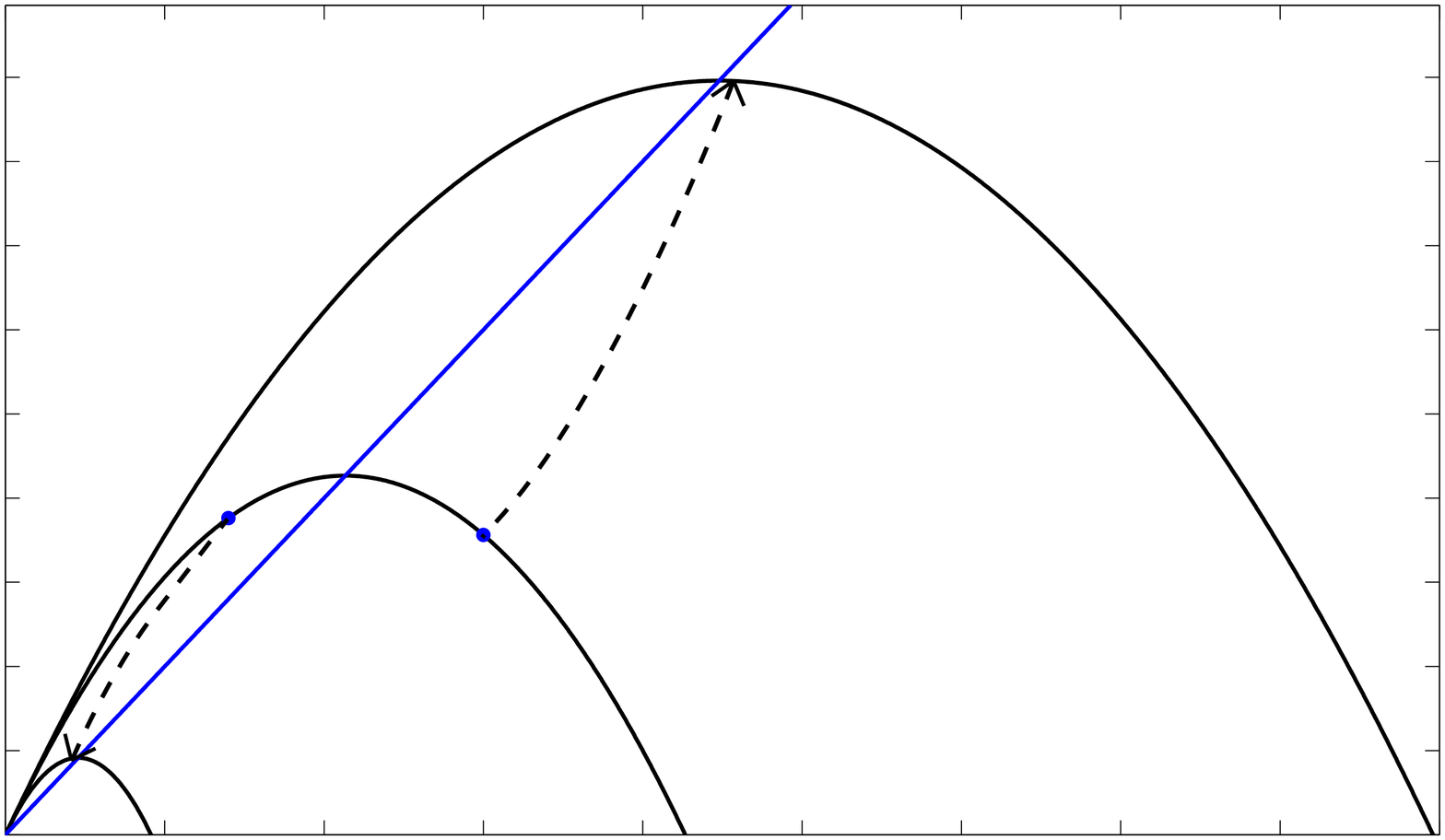}
}
\caption{\label{fig02} $(a)$ An example curve for the function $C(\phi)$ 
with a minimum solids volume fraction $\phi_0$ and a vertical asymptote at $\phi_0+O(\eps)$. 
$(b)$ Nested yield surfaces of the form \eqref{Cmu-ansatz-for-aux-fcn} for a fixed value of $I$  
with differing solids volume fractions. 
(The solid blue line, the dashed arrows, and the labels
$A$ and $B$ refer to a discussion of CSSM in the Appendix~1.)}
\end{center}
\end{figure}

Without making any attempt to be general, we illustrate one example of how $\mu(I)$-rheology
may be included in constitutive relations of the form \eqref{yield-cond-Cmu}, \eqref{flow-rule-Cmu}.
Motivated by equation \eqref{sample-yield-cond} in Appendix 1, we make the ansatz
\begin{equation} \label{Cmu-ansatz-for-aux-fcn}
\begin{array}{cccl}
 \mbox{(a)}& Y(p,\phi,I)& =& \alpha(I)p-p^2/C(\phi) \\ \\
 \mbox{(b)}& f(p,\phi,I)& =& \beta(I)-2p/C(\phi) \, .
\end{array}
\end{equation}
In these equations, it is worth emphasising that $p,\phi, I$ are treated as independent variables, not to be confused with the dependence of $I$ on $p$ in the previous subsection. 
The function $C(\phi)$ is an increasing function of $\phi$.   As $\phi$ varies (with $I$ fixed) the yield loci 
$ \|\taub\| = Y(p,\phi,I)$ derived from (\ref{Cmu-ansatz-for-aux-fcn}a) 
form a nested family of convex curves in stress space (cf. Figure~\ref{fig02}(b)).
Observe from \eqref{flow-rule-Cmu} that deformation without volumetric strain 
is possible if $f(p,\phi,I)=0$; i.e., for (\ref{Cmu-ansatz-for-aux-fcn}b), if $p/C(\phi)=\beta(I)/2$.
Substituting this formula into \eqref{yield-cond-Cmu} and using (\ref{Cmu-ansatz-for-aux-fcn}a), we derive
$$
  \|\taub\|= [\alpha(I)-\beta(I)/2]\; p 
$$
for such isochoric deformation to be possible.
Thus, to recover the yield condition $\|\taub\|= \mu(I)p$ of the $\mu(I)$-rheology,
let us require that
\begin{equation} \label{Cmu-limit-beh-2}
 \alpha(I)-\beta(I)/2=\mu(I) \, .
\end{equation}

\bigskip

\noindent{\bf Lemma 1.} Equations \eqref{Cmu-well-posed-cond} and \eqref{Cmu-limit-beh-2}
imply that
\begin{equation} \label{Cmu-formulas-for-alpha}
 \alpha(I) = \frac{4}{5}\mu(I) +\frac{12}{25} I^{-2/5} \int_0^I \tilde{I}^{-3/5} \mu(\tilde{I}) d\tilde{I}
\end{equation}
and
\begin{equation} \label{Cmu-formulas-for-beta}
 \beta(I) = -\frac{2}{5}\mu(I) +\frac{24}{25} I^{-2/5} \int_0^I \tilde{I}^{-3/5} \mu(\tilde{I}) d\tilde{I} \, .
\end{equation}.

\bigskip

\noindent {\bf Proof:} \
Substituting the relations \eqref{Cmu-ansatz-for-aux-fcn} into \eqref{Cmu-well-posed-cond}, and using \eqref{Cmu-limit-beh-2} to eliminate $\beta,$ we derive the linear ordinary differential equation for $\alpha=\alpha(I):$
\begin{equation}
\frac 52 I\alpha'(I)+\alpha(I)=2\mu(I)+2I\mu'(I)\, .
\end{equation}
Solving this linear equation for $\alpha(I),$ with an integrating factor,
we obtain
$$
I^{2/5} \alpha(I)=\frac 45 \int_0^I \tilde{I}^{-3/5}\mu(\tilde{I})d\tilde{I} +\frac 45 \int_0^I \tilde{I}^{2/5}\mu'(\tilde{I})d\tilde{I} \, ,
$$
from which 
  the formula \eqref{Cmu-formulas-for-alpha} follows after integrating the second integral by parts.
  Finally, substituting this formula for $\alpha(I)$ into \eqref{Cmu-limit-beh-2}, we obtain the formula \eqref{Cmu-formulas-for-beta}  for $\beta(I).$ \proofend

\bigskip

\noindent{\bf Lemma 2.} The yield condition and flow-rule function (\ref{Cmu-ansatz-for-aux-fcn}a,b) that follow from
\eqref{Cmu-formulas-for-alpha}, \eqref{Cmu-formulas-for-beta} verify hypotheses
\eqref{Cmu-well-posed-cond} and \eqref{ineq-for-thm}, provided $\mu(I)$ satisfies \eqref{cond-on-mu}.

\bigskip

\noindent {\bf Proof:} \ Of course \eqref{Cmu-well-posed-cond} is satisfied because this equation was imposed in deriving 
\eqref{Cmu-formulas-for-alpha}, \eqref{Cmu-formulas-for-beta}.

 Differentiating  (\ref{Cmu-ansatz-for-aux-fcn}b), we see that
$\partial_p f(p,\phi,I)=-2/C(\phi)<0.$ To calculate $\partial_I f(p,\phi,I),$ we first 
reparametrize the integral in \eqref{Cmu-formulas-for-beta} to obtain
$\beta(I)=-\frac 25 \mu(I)+\frac{24}{25}    \int_0^1 s^{-3/5}\mu(sI)ds \,.$ Then
$$
\partial_I f(p,\phi,I)=\beta'(I)=-\frac 25 \mu'(I)+\frac{24}{25} \int_0^1 s^{2/5}\mu'(sI)ds \, .
$$
By \eqref{cond-on-mu}, $\, \mu''(I)<0 $, so
$\mu'(sI)>\mu'(I)$ for $0<s<1.$ Thus,
$$
\beta'(I)>\mu'(I)\left\{-\frac 25+\frac{24}{25} \int_0^1 s^{2/5}ds\right\}
=\mu'(I)\left\{\frac{24}{35}-\frac 25\right\}>0 \, ,
$$
the last inequality using \eqref{cond-on-mu}.
Consequently, 
\begin{equation}\label{f2}
\partial_p f(p,\phi,I)-\frac{I}{2p}\partial_I f(p,\phi,I)<0 \, ,
\end{equation}
proving inequality (\ref{ineq-for-thm}b).

For inequality (\ref{ineq-for-thm}a), we reparametrize the integral \eqref{Cmu-formulas-for-alpha} 
and differentiate to obtain
$$
\partial_I Y(p,\phi,I)=\alpha'(I)p=p\left(\frac 45 \mu'(I)+\frac{12}{25}    \int_0^1 s^{2/5}\mu'(sI)ds\right)>0 \, 
,
$$
as desired. \proofend

\bigskip

Based on an analogy with CSSM, let us suppose that $C(\phi)$ is a sensitive function of $\phi$, say of the form
\begin{equation} \label{form-of-C}
 C(\phi)=\tilde{C}\left(\frac{\phi-\phi_0}{\eps} \right)
\end{equation}
where $\phi_0$ is the minimum solids fraction for sustained stress transmission between grains
(random loose packing),  $\eps$ is a
small parameter, and for definiteness we may take $\tilde{C}(z)=z/(1-z)$ as in figure \ref{fig02}.
Note that $C(\phi)$ diverges as $\phi\to\phi_0+\eps$; thus,
\eqref{form-of-C} requires that $\phi$ is confined to a narrow range,
\begin{equation} \label{range-phi}
    \phi_0 \le \phi < \phi_0+\eps \, .
\end{equation}
In physical terms, the maximum solids fraction $\phi_0+\eps$ represents the jamming threshold.
We call the limit $\eps\to0$ \emph{incompressible} because, as may be seen from
\eqref{range-phi}, the density of material becomes essentially constant.

\bigskip

\noindent{\bf Lemma 3.} As $\eps\to0$, the CIDR equations
reduce to the equations of incompressible $\mu(I)$-rheology, \eqref{incompress}, \eqref{mu(I)-system-mom}.

\bigskip

\noindent{\bf Proof:} \ 
We process the CIDR equations, which have the six unknowns $\phi$, $u_i$, and $\sigma_{ij}$, as follows.
First we reduce to five unknowns---$\phi$, $u_i$, $p$ and $\tau=\|\taub\|$---by recalling the definition
 \eqref{dfn-tau}
and the alignment condition \eqref{align-coulomb} to write
$$
  \sigma_{ij} = -p \delta_{ij} + \tau \frac{D_{ij}}{\|\u{D}\|} \, .
$$
Next we use the yield condition to eliminate $\phi$, reducing this number to four.  Specifically,
substituting (\ref{Cmu-ansatz-for-aux-fcn}a) into \eqref{yield-cond-Cmu}, we write the yield condition
\begin{equation} \label{new-eqn-10-10}
 \tau=\alpha(I) p -p^2/C(\phi) \, .
\end{equation}
Solving \eqref{new-eqn-10-10} for $\phi$ we obtain
\begin{equation} \label{phi-of-tau}
 \phi = \Phi(\nabla \u{u},p,\tau)=C^{-1}\left(\frac{p^2}{\alpha(I) p-\tau} \right) \, ,
\end{equation}
where the dependence on $\nabla \u{u}$ comes from the fact that
$I=2d\|\u{D}\|/\sqrt{p/\rho_*}.$ 
Substitution of this formula into the conservation laws \eqref{c-mass}, \eqref{c-mom} yields the equations
\begin{subequations} \label{eqn-motion-CSSM}
  \begin{align}
	(\partial_t +u_j \partial_j) \Phi(\nabla \u{u},p,\tau) + \Phi(\nabla \u{u},p,\tau)\, \mbox{div } \u{u}
  &=   0 \, , \label{eqn-motion-CSSMa} \\
 \rho_*\Phi(\nabla \u{u},p,\tau)(\partial_t  +u_j \partial_j) u_i  &=  
 \partial_j  \left[\frac{\tau}{\|\u{D}\|} D_{ij} \right]
-\partial_i p +\rho_*\Phi g_i \, . \label{eqn-motion-CSSMb} 
  \end{align}
\end{subequations}
Finally, we show the flow rule \eqref{flow-rule-Cmu} may be rewritten
\begin{equation} \label{flow-rule-form-for-limit}
  \mbox{div }\u{u} = 4[\tau/p-\mu(I)] \, \|\Db\| \, .
\end{equation}
To see this, we combine (\ref{Cmu-ansatz-for-aux-fcn}b) with \eqref{Cmu-limit-beh-2} to conclude
$$
  f(p,\phi,I)=\beta(I) -2p/C(\phi) = 2[\alpha(I)-\mu(I)] -2p/C(\phi)
$$
and substitute the relation $\alpha(I)=\tau/p+p/C(\phi)$ derived by manipulating \eqref{new-eqn-10-10}.
Thus, the system \eqref{eqn-motion-CSSM}, \eqref{flow-rule-form-for-limit}
governs the evolution of the four unknowns $u_i$, $p$, and $\tau$.

Now we claim that if $C(\phi)$ has the form \eqref{form-of-C}, then \eqref{eqn-motion-CSSM}, 
\eqref{flow-rule-form-for-limit} is a singular perturbation of \eqref{incompress}, \eqref{mu(I)-system-mom}. It 
follows from \eqref{form-of-C} that \eqref{phi-of-tau} has the expansion
\begin{equation} \label{phi-of-tau-expn}
 \phi=\phi_0+\eps\tilde{\Phi}(\nabla \u{u},p,\tau) \, ,
\end{equation}
where $\tilde{\Phi}(\nabla \u{u},p,\tau)=\tilde{C}^{-1}(p^2/[\alpha(I) p-\tau])$. Substituting 
\eqref{phi-of-tau-expn} into the continuity equation (\ref{eqn-motion-CSSM}a) we find
$$
\eps(\partial_t +u_j \partial_j) \tilde{\Phi}(\nabla \u{u},p,\tau)
+ [\phi_0 + \eps\tilde{\Phi}(\nabla \u{u},p,\tau)]\, \mbox{div } \u{u} = 0 \, .
$$
If $\eps=0$ then this equation reduces to $\mbox{div }\u{u}=0$, although this is of course a highly singular limit. 
Thus, if $\eps=0$, the left hand side of \eqref{flow-rule-form-for-limit} vanishes, 
so this equation simplifies to the yield condition $\tau=\mu(I) p$, and substitution into 
\eqref{eqn-motion-CSSMb} yields \eqref{mu(I)-system-mom}. This proves the lemma. \proofend

\section{Proofs, Part I: Linearization}

\subsection{An alternative formulation of the alignment condition}

It is convenient to study the linearized equations with a reformulated alignment condition that
describes stress in terms of eigenvectors of,
rather than entries of, the stress tensor. 
Since $\u{\tau}$ defined by \eqref{dfn-tau} has trace zero, it has
eigenvalues%
\footnote{Hence $\sigmab$ has eigenvalues $-p\pm\|\taub\|$.  Note that $-p-\|\taub\|$ is
the major stress eigenvalue---although this eigenvalue is the smaller algebraically,
it is the larger in absolute value.}
 $\pm\|\taub\|$. Taking $\psi$ as the angle that the eigenvector with eigenvalue $-\|\taub\|$ makes with the $x_1$-axis gives
\begin{equation} \label{interpret-psi}
 \u{\tau} = -\|\u{\tau}\| \btwo \cos2\psi& \sin2\psi\\ \sin2\psi& -\cos2\psi \etwo \, ,
\end{equation}
which may be verified by checking that $(\cos\psi,\sin\psi)$ is an eigenvector of this
matrix with eigenvalue $-\|\taub\|$. Thus, the stress tensor $\sigma_{ij}$ is completely specified by the three scalars $p$, $\|\u{\tau}\|$, and $\psi$.

Focusing on the first rows of the strain-rate tensor \eqref{dev-strain-rate} and of \eqref{interpret-psi}, we
extract from the matrix equation \eqref{align-coulomb} the vector equation
\begin{equation} \label{align-first-row}
  (\partial_1 u_1-\partial_2 u_2, \partial_1 u_2+\partial_2 u_1) = k(\cos2\psi,\sin2\psi) \, ,
\end{equation}
where $k=-2\|\u{D}\|<0$. Since $\u{D}$ and $\u{\tau}$ lie in the two-dimensional space of trace-free,
symmetric matrices, \eqref{align-first-row} is equivalent to \eqref{align-coulomb}. It follows from \eqref{align-first-row} that
\begin{equation} \label{align-CSSM}
\mbox{\bf Alt. alignment:} \qquad
 (\partial_1 u_2 + \partial_2 u_1) \cos2\psi - (\partial_1 u_1 -\partial_2 u_2) \sin2\psi =0 \, .
\end{equation}
In point of fact, this equation is slightly weaker than the alignment condition since \eqref{align-CSSM} is consistent with the possibility that $k>0$ in \eqref{align-first-row}; to rule out the latter
possibility we impose the supplemental inequality%
\footnote{It is also true that $(\partial_1 u_2 + \partial_2 u_1)\sin2\psi\le0$, and if
$\cos2\psi$ were to vanish, we would need to use this inequality to guarantee that $k<0$.
However, this issue will not arise in the analysis below.}
 that
\begin{equation} \label{align-ineq}
 (\partial_1 u_1 -\partial_2 u_2)\cos2\psi\le0 \, .
\end{equation}
%

\subsection{The calculation}

Substitution of the stress tensor \eqref{interpret-psi} into the momentum balance equations \eqref{c-mom} allows
for the full set of equations to be written as
\begin{subequations}
  \begin{align}
    \rho_* \phi \left(\d{t} + u_1 \d{1}  +u_2 \d{2} \right)u_1+\d{1}\left[p+\tau \cos(2\psi)\right]
+\d{2}\left[\tau \sin(2\psi)\right]&=\rho_*\phi g_1 \, , \label{12a} \\
    \rho_* \phi \left(\d{t} + u_1 \d{1}  +u_2 \d{2} \right)u_2
+\d{1}\left[\tau \sin(2\psi)\right]+\d{2}\left[p-\tau \cos(2\psi)\right]&=\rho_* \phi g_2 \, , \label{12b} \\
 (\d{t}+ u_1 \d{1}  +u_2 \d{2})\phi+ \phi\left( \d{1} u_1 + \d{2}u_2\right) &=0 \, , \label{12c}  \\
\d{1} u_1 + \d{2}u_2 &= 2 f \| \u{D}\| \, , \label{12d}  \\
\left(\d{2} u_1 + \d{1}u_2\right)\cos(2\psi)+\left(\d{2} u_2 - \d{1}u_1\right)\sin(2\psi) &=
0 \, . \label{12e}
  \end{align}
\end{subequations}
This system has five scalar unknowns, $\u{U}=(u_1,u_2,\phi,p,\psi)$.
In \eqref{12a},\eqref{12b}, $\,\tau$ is a mnemonically suggestive abbreviation for the yield
function $Y(p,\phi,I)$ in \eqref{yield-cond-Cmu}, and in \eqref{12d}, a repetition of
\eqref{flow-rule-Cmu}, the function $f$ depends on arguments $(p,\phi,I)$
that are not written explicitly.

As in Appendix~2, to linearise the equations we substitute a perturbation of a base solution
$\u{U}^{(0)}(\xb,t)$, say
\begin{equation}\label{e13}
\u{U} = \u{U}^{(0)} + \u{\hat{U}}  \, ,
\end{equation}
into the equations, retain only terms that are linear in the perturbation $\u{\hat{U}}$,
and freeze the coefficients at an arbitrary point $(\xb^*,t^*)$.
It is convenient to temporarily drop most terms not of maximal order and estimate their effect in
a calculation at the end of the argument.
For example, this construction applied to \eqref{12c} yields the the constant-coefficient, linear equation
\begin{equation} \label{12c-lin}
  (\d{t}+ u_1^*\, \d{1}  + u_2^*\, \d{2}) \hat{\phi}
+ \phi^* (\d{1} \hat{u}_1 +\d{2} \hat{u}_2) =0
\end{equation}
where $u_{j}^* =u\szero_j(\xb^*,t^*)$ and $\phi^*=\phi\szero(\xb^*,t^*)$.
Lower-order terms $\d{j} \phi^*\, \hat{u_j}$ and $\d{j}u_j^* \, \hat{\phi}$
in the full linearisation of \eqref{12c} have been dropped in \eqref{12c-lin}.

\begin{figure}
\begin{center}
\SetLabels
       \endSetLabels
       \strut\AffixLabels{ }
\SetLabels
      \L (0.02*0.5) $x_2$\\
      \L (0.5*0.0) $x_1$\\
       \endSetLabels
       \strut\AffixLabels{
	\includegraphics[width=0.6\textwidth]{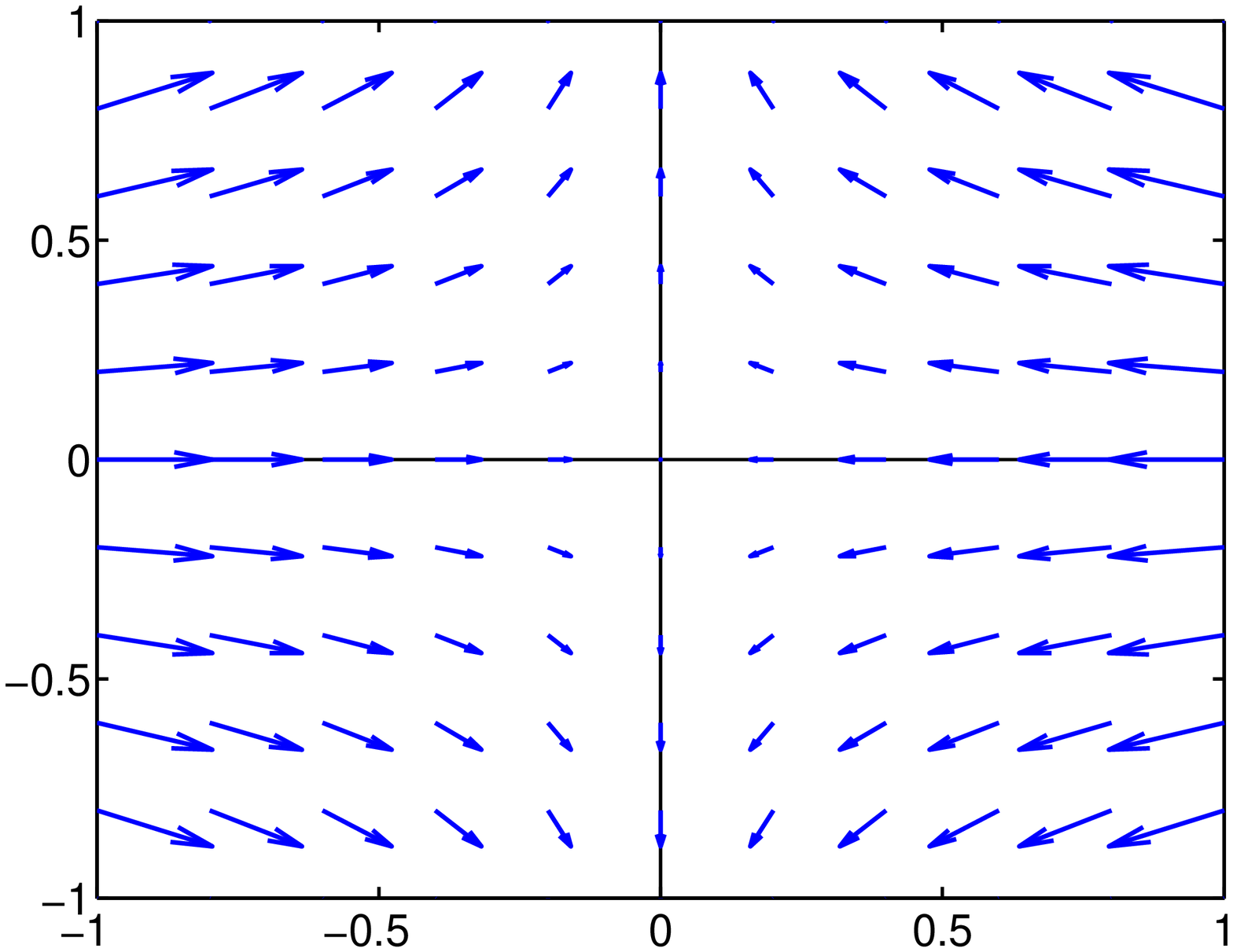}
	}

\caption{\label{fig03} An example of a base-state velocity field for the strain-rate tensor \eqref{e14}
with $\d{1}u_1^{(0)}\equiv-1$ and $\d{2}u_2^{(0)}\equiv1/2$. }
\end{center}
\end{figure}

In expanding the fully nonlinear
factor $\|\u{D}\|$ in \eqref{12d},
we may take advantage of the rotational invariance of the equations to arrange that
$\psi^*=0$; i.e., we may calculate in a rotated coordinate system for which, at $(\xb^*,t^*)$,
the $x_1$-axis is the maximal
stress axis. Then by the alignment condition \eqref{align-CSSM} the base-state deviatoric strain-rate
tensor is diagonal at $(\xb^*,t^*)$
\begin{equation}\label{e14}
\u{D}^* = \btwo \left(\d{1}u_1^* -  \d{2}u_2^*\right)/2 & 0 \\
    0 & \left(\d{2}u_2^*-\d{1}u_1^*\right)/2 \etwo  \, ,
\end{equation}
and by \eqref{align-ineq}, in the 1,1-position of this matrix $\d{1}u_1^* -  \d{2}u_2^*<0$.
This corresponds to non-zero compression along the major stress axis,
as illustrated in Figure \ref{fig03}.
Now
\begin{subequations}
  \begin{align}
\|  (\u{D}^*+\u{\hat{D}}) \| & = \frac{1}{2}
\left[ (\d{1}u_1^*-\d{2}u_2^*+\d{1}\hat{u}_1-\d{2}\hat{u}_2)^2
+(\d{2}\hat{u}_1+\d{1}\hat{u}_2)^2 \right]^{1/2} \,  \label{15a}  \\
& \approx \|  \u{D}^* \| -  \left( \d{1}\hat{u}_1-\d{2}\hat{u}_2\right)/2 \, , \label{15b} &
  \end{align}
\end{subequations}
\vspace*{-0.1mm}where the approximation follows from the expansion
$$
\sqrt{(-A+X)^2+Y^2}=A-X+\Oh\left(X^2+Y^2\right)
$$
if $A>0$ and $|X|,|Y|\ll A\,$.
Thus,  as given in Table~\ref{table-compute-lin}, 
the (local) linearisation of $\|\u{D}\|$ equals
$  -\left( \d{1}\hat{u}_1-\d{2}\hat{u}_2\right)/2\,$.

\begin{table}

\begin{center}
\begin{tabular}{cc}
Term in \eqref{12a}-\eqref{12e}& Contribution to \eqref{21a}-\eqref{21e}	 \\
\hline
\\
$\|\u{D}\|$&	$-\hat{D}_{11}$\\
\\
$I$&		$-\frac{I^*}{\|\u{D}^*\|}\hat{D}_{11} -\frac{I^*}{2p^*}\hat{p}$\\
\\
$\partial_j[\tau\cos(2\psi)]$&	
    $(\partial_p \tau)^{\!*} \, \partial_j \hat{p}
    +(\partial_\phi \tau)^{\!*} \partial_j \hat{\phi}$ \\
&	$+(\partial_I \tau)^{\!*}
\left\{ -\frac{I^*}{\|\u{D}^*\|}\partial_j\hat{D}_{11}
	    -\frac{I^*}{2p^*} \partial_j \hat{p} \right\}$\\
\\
$\partial_j[\tau\sin(2\psi)]$& $2\tau^* \partial_j \hat{\psi}$ \\
\\
$f\|\u{D}\|$& $-f^*\hat{D}_{11} +
    \|\u{D}^*\|(\partial_p f)^{\!*} \hat{p}
    +\|\u{D}^*\| (\partial_\phi f)^{\!*} \hat{\phi}$ \\
&	$+\|\u{D}^*\|(\partial_I f)^{\!*}
\left\{ -\frac{I^*}{\|\u{D}^*\|}\hat{D}_{11}
	    -\frac{I^*}{2p^*}  \hat{p} \right\}$\\

\end{tabular}
\end{center}
\caption{\em List of maximal-order linearisations of terms in \eqref{12a}-\eqref{12e}, to assist
in deriving \eqref{21a}-\eqref{21e}.  In this table only, the abbreviation
$\hat{D}_{11}= (\d{1}\hat{u}_1-\d{2}\hat{u}_2)/2$ is used.}
\label{table-compute-lin}

\end{table}

In \eqref{12d} the function $f$ contains $p$, $\phi$, and $I$ as implicit arguments.
As reflected in the table, the dependence on $p$ and $\phi$ contributes zeroth-order terms in these variables
to the linearisation.

In \eqref{12a}, \eqref{12b}, $\,\tau$ also depends on $p$, $\phi$, and $I$, and the
terms involving $\tau$ are differentiated; hence new issues arise in linearising them.
For example, by the chain rule,
$$ \begin{array}{c}
  \d{j}[\tau \cos(2\psi)]=  \cos(2\psi)
    \left\{\partial_p\tau\, \d{j} p +\partial_\phi\tau\, \d{j}\phi 
+ \partial_I \tau \,\left[\frac{2d}{\sqrt{p/\rho_*}}\d{j} \|\u{D}\| 
      -\frac{d\|\u{D}\|}{\sqrt{p^3/\rho_*}} \d{j} p \right] \right\} \\
-2\tau\sin(2\psi) \d{j}\psi  \,.
\end{array}
$$
Since $\psi^*=0$, the full linearisation of, say, the first term here equals 
$(\partial_p \tau)^{\!*} \, \d{j} \hat{p}$, a term given in the table, 
plus lower-order terms
$$
  (\d{j}p)^*\left\{(\partial_{pp}\tau)^* \, \hat{p} +(\partial_{\phi p}\tau)^*\, \hat{\phi}
      +(\partial_{Ip}\tau)^* \, \left[-\frac{I^*}{\|\Db^*\|}\hat{D}_{11} 
-\frac{I^*}{2p^*}\hat{p}  \right] \right\}\,.
$$
All of these terms, as well as numerous other analogous terms 
in the full linearisation of \eqref{12a} that are not of maximal order, have been 
dropped in \eqref{21a}-\eqref{21e}.

Putting all the pieces together, we obtain the linearisation%
\footnote{These equations are maximal order except that in \eqref{21a} and \eqref{21b} the term
$d_t^*\hat{u}_j$ retains first-order spatial derivatives even though these equations also contain second-order
derivatives of $\hat{u}_j$.}
of the system \eqref{12a}-\eqref{12e}
\begin{subequations}
  \begin{align}
	\rho_* \phi^* d_t^*\hat{u}_1+ A\left(-\d{{11}}\hat{u}_1+\d{{12}}\hat{u}_2\right)
+(\d{\phi}\tau)^{\!*}\d{1}\hat{\phi}+\left(1+B\right)\d{1}\hat{p}+2\tau^*\d{2}\hat{\psi} &= 0\, , \label{21a}
\\
	\rho_* \phi^* d_t^*\hat{u}_2+A\left(\d{{12}}\hat{u}_1-\d{{22}}\hat{u}_2\right)
-(\d{\phi}\tau)^{\!*}\d{2}\hat{\phi}+\left(1-B\right)\d{2}\hat{p}+2\tau^*\d{1}\hat{\psi} &= 0 \, , \label{21b}
 \\
d_t^*\hat{\phi}+\phi^*(\d{1}\hat{u}_1  +\d{2}\hat{u}_2 )  &= 0 \, , \label{21c}  \\
\left(1+C\right)\d{1} \hat{u}_1 + \left(1-C\right)\d{2}\hat{u}_2
-2\| \u{D^*}\| (\d{\phi}f)^{\!*} \hat{\phi} +\Gamma \hat{p}&=0 \, , \label{21d}  \\
\d{2}\hat{u}_1+\d{1}\hat{u}_2+4\| \u{D}^*\|\hat{\psi} &= 0 \, , \ \label{21e}
  \end{align}
\end{subequations}
where
\begin{equation}\label{e22-a}
	d_t^*=\d{t}+u_1^*\d{1}+u_2^*\d{2} \, , \;\;\;
	A = \frac{I^*}{2\| \u{D}^*\|}(\d{I}\tau)^{\!*} \, , \;\;\;
	B = (\d{p}\tau)^{\!*}-\frac{I^*}{2p^*}(\d{I}\tau)^{\!*} , \;\;\;
\end{equation}
\begin{equation} \label{e22-b}
	C = f^* + I^* (\d{I}f)^{\!*}\,, \;\;\; \mbox{and} \;\;\;
	\Gamma =  -2\| \u{D^*}\| \left( (\d{p}f)^{\!*}-\frac{I^*}{2p^*}(\d{I}f)^{\!*}\right).
\end{equation}

\noindent Observe that by hypothesis \eqref{Cmu-well-posed-cond}, $\;B=C$, a fact that we use in \eqref{e26} and 
below.

\section{Proofs, Part II: Calculation of growth rates}
\subsection{The eigenvalue problem}

We now look for exponential solutions of
\eqref{21a}-\eqref{21e},
\begin{equation}\label{e23}
\u{\hat{U}}(\xb,t) = e^{i\langle\xib,\xb\rangle+\lambda t} \u{\tilde{U}}\, ,
\end{equation}
where $\u{\tilde{U}}=(\tilde{u}_1,\tilde{u}_2,\tilde{\phi},\tilde{p},\tilde{\psi})$ is a 5-vector of scalars,
$\xib=(\xi_1,\xi_2)$ is a vector wavenumber, $\langle\,,\rangle$ indicates the inner product, and 
$\lambda$ is the growth rate.
The function \eqref{e23} is a solution of \eqref{21a}-\eqref{21e} iff $\lambda, \u{\tilde{U}}$ satisfies the
generalised eigenvalue problem
\begin{equation}\label{e24}
\u{S} \u{\tilde{U}} = -(\lambda+i\langle\u{u}^*,\xib\rangle )\u{E}\u{\tilde{U}} \, ,
\end{equation}
where $\u{u}^*=(u_1^*,u_2^*)$,
\begin{equation}\label{e26}
\u{S} = \bfive A\xi_1^2 & -A\xi_1\xi_2 & i(\d{\phi}\tau)^{\!*}\xi_1 & (1+B)i\xi_1 & 2i\tau^*\xi_2 \\
		-A\xi_1\xi_2 & A\xi_2^2 & -i(\d{\phi}\tau)^{\!*}\xi_2 & (1-B)i\xi_2 & 2i\tau^*\xi_1 \\
		 i\phi^*\xi_1 & i\phi^*\xi_2 & 0 & 0 & 0 \\
		  (1+B)i\xi_1 & (1-B)i\xi_2 & -2\| \u{D^*}\|(\d{\phi}f)^{\!*} & \Gamma & 0 \\
		   i\xi_2 & i\xi_1 & 0 & 0 & 4\|\u{D}^*\|  \efive  \, ,
\end{equation}
and
\begin{equation}\label{e25}
\u{E} = \bfive \rho_*\phi^* &  &  &  & \\
		 & \rho_*\phi^* &  &  &  \\
		  &  & 1 &  &  \\
		   &  &  & 0 &  \\
		    &  &  &  & 0  \efive  \, .
\end{equation}
On the right side of \eqref{e24}, the modified eigenvalue parameter is
$\lambda+i\langle\u{u}^*,\xib\rangle$ because
$$
    d_t^* e^{i\langle\xib,\xb\rangle+\lambda t} =
(\lambda+i\langle\u{u}^*,\xib\rangle )
e^{i\langle\xib,\xb\rangle+\lambda t} .
$$

Equation \eqref{e24} is a \emph{generalised} eigenvalue problem because $\u{E}$, the matrix of coefficients of
time-derivative terms, is not invertible.
To extract an ordinary eigenvalue problem, we decompose $\u{S}$ into blocks
\begin{equation}\label{e29}
\u{S} = \btwo \u{S}_{11} & \u{S}_{12} \\
		   \u{S}_{21} & \u{S}_{22}  \etwo  \, ,
\end{equation}
where
\begin{equation}\label{e30}
\u{S}_{11} = \bthree A\xi_1^2 & -A\xi_1\xi_2 & i(\d{\phi}\tau)^{\!*}\xi_1  \\
		-A\xi_1\xi_2 & A\xi_2^2 & -i(\d{\phi}\tau)^{\!*}\xi_2 \\
		  i\phi^*\xi_1 & i\phi^*\xi_2 &0 \ethree  \,
\end{equation}
and $\u{S}_{12}$, $\u{S}_{21}$, and $\u{S}_{22}$ fill out the rest of the matrix.
Defining $\u{\tilde{U}}_1=(\tilde{u}_1,\tilde{u}_2,\tilde{\phi})$ and
$\u{\tilde{U}}_2=(\tilde{p},\tilde{\psi})$, we rewrite \eqref{e24} as
\begin{equation}\label{e31}
 \btwo \u{S}_{11} & \u{S}_{12} \\ \u{S}_{21} & \u{S}_{22}  \etwo   \bvec \u{\tilde{U}}_1  \\ \u{\tilde{U}}_2   \evec
=- (\lambda+i\langle\u{u}^*,\xib\rangle )  \u{E}
\bvec \u{\tilde{U}}_1  \\ \u{\tilde{U}}_2   \evec  \, .
\end{equation}
The zero entries in the last two rows of $\u{E}$  mean that $\u{S}_{21}\u{\tilde{U}}_1 + \u{S}_{22}\u{\tilde{U}}_2=0$
so we can solve for
\begin{equation}\label{e32}
\u{\tilde{U}}_2=-\u{S}_{22}^{-1}\u{S}_{21}\u{\tilde{U}}_1  \, .
\end{equation}
Substitution of $\tilde{\u{U}}_2$ into \eqref{e31} then reduces this problem%
\footnote{In other words, we are performing on the symbol level the reduction that we performed on the
operator level in Section~2.4.}
 to the ordinary $3\times3$ eigenvalue problem,
\begin{equation}\label{e33}
  \u{E}_{11}^{-1} \left[\u{S}_{11}-\u{S}_{12}\u{S}_{22}^{-1}\u{S}_{21}\right]\u{\tilde{U}}_1
=  -(\lambda +i\langle\u{u}^*,\xib\rangle ) \u{\tilde{U}}_1   \,
\end{equation}
where $\u{E}_{11}$ is the $3\times3$ block in the upper left of $\u{E}$.

We decompose the $3\times3$ matrix in \eqref{e33} into smaller blocks,
\begin{equation}\label{e34}
\btwo (\u{M}+\u{N})/\rho_* \phi^* & i\u{V}/\rho_* \phi^* \\ i\phi^* \u{\xi}^T & 0
\etwo\u{\tilde{U}}_1 = -(\lambda +i\langle\u{u}^*,\xib\rangle )  \u{\tilde{U}}_1  \,
\end{equation}
where we calculate
\begin{equation}\label{e39}
\u{M}= A \btwo \xi_1^2  & -\xi_1\xi_2 \\
		-\xi_1\xi_2 & \xi_2^2  \etwo \,
\end{equation}
as the contribution of $\u{S}_{11}$,
\begin{equation}\label{new-formula-N}
\u{N} =  \btwo \frac{(1+B)^2}{\Gamma }\xi_1^2 +\frac{\tau^*}{2\|\u{D}^*\| }\xi_2^2 &
\frac{(1-B^2)}{\Gamma }\xi_1\xi_2 +\frac{\tau^*}{2\|\u{D}^*\| }\xi_1\xi_2 \\
		\frac{(1-B^2)}{\Gamma }\xi_1\xi_2 +\frac{\tau^*}{2\|\u{D}^*\| }\xi_1\xi_2 &
\frac{(1-B)^2}{\Gamma }\xi_2^2 +\frac{\tau^*}{2\|\u{D}^*\| }\xi_1^2   \etwo  \, ,
\end{equation}
as the contribution of $-\u{S}_{12}\u{S}_{22}^{-1}\u{S}_{21} $, which is symmetric,
and
\begin{equation}\label{e36}
\u{V} =  \bvec \left((\d{\phi}\tau)^{\!*}+\frac{2(1+B) \| \u{D^*}\|(\d{\phi}f)^{\!*}}{\Gamma} \right)\xi_1 \\
\\
\left(-(\d{\phi}\tau)^{\!*}
+ \frac{2(1-B) \| \u{D^*}\|(\d{\phi}f)^{\!*}}{\Gamma} \right)\xi_2 \evec  \, .
\end{equation}

\subsection{Estimation of the eigenvalues}

We claim that the growth-rate eigenvalues
 \eqref{e34} satisfy
$$
  \max_{j=1,2,3} \quad \sup_{\xib\in\R^2} \quad \Re \lambda_j(\xib) < \infty \, .
$$
By compactness, it suffices to prove that
\begin{equation} \label{wellp-first-formula}
  \max_{j=1,2,3} \quad \limsup_{|\xib|\to\infty} \quad \Re \lambda_j(\xib) < \infty \, .
\end{equation}
Since only the real parts of eigenvalues matter, we may drop the term
$i\langle\u{u}^*,\xib\rangle$ in \eqref{e34} and verify \eqref{wellp-first-formula} for the eigenvalue problem%
\footnote{Don't forget the minus sign in this equation---the growth rates are \emph{negative}
eigenvalues of $\u{P}$.}
\begin{equation} \label{wellp-second-formula}
  \u{P} \u{\tilde{U}} = -\lambda \u{\tilde{U}}
\end{equation}
where we write
\begin{equation} \label{matrix}
 \u{P} = \btwo (\u{M}+\u{N})/\rho_* \phi^* & i\u{V}/\rho_* \phi^* \\ i\phi^* \u{\xi}^T & 0 \etwo
\end{equation}
for the matrix in \eqref{e34}
and we shorten the notation by dropping the subscript 1 on $\u{\tilde{U}}$.
For large $\xib$ it is instructive to use perturbation theory to
compare the eigenvalues  \eqref{wellp-second-formula} with the eigenvalues
$\u{P}_0  \u{\tilde{U}} =-\Lambda \u{\tilde{U}}$ where
\begin{equation} \label{evalue-comparison}
  \u{P}_0 =(\rho_* \phi^*)^{-1} \btwo \u{M}+\u{N} & 0 \\ 0& 0 \etwo.
\end{equation}

\bigskip

\noindent{\bf Lemma. }\emph{Provided } $\xib\neq0$, \emph{ the $2\times2$ matrix } $\u{M}+\u{N}$ \emph{ is
positive definite.}

\bigskip

\noindent{\bf Proof.} Since $\u{M}$ and $\u{N}$ are symmetric, it suffices to show that the trace and determinant
of $\u{M}+\u{N}$ are positive.
According to \eqref{ineq-for-thm}, $\;A>0$ and $\Gamma>0$, from which
it follows immediately that
$\mbox{tr } (\u{M}+\u{N})>0$.

Regarding the determinant, for any $2\times2$ matrices
\begin{equation}\label{e41}
\det(\u{M}+\u{N})=\det \u{M}+\det \u{N} + \chi(\u{M},\u{N}) \, ,
\end{equation}
where
\begin{equation}
\chi(\u{M},\u{N}) =  M_{22} N_{11} + M_{11}N_{22} - M_{12}N_{21}-M_{21}N_{12}\,  \label{42a}
\end{equation}
accounts for the cross terms.
For the specific matrices \eqref{e39} and \eqref{new-formula-N}, $\,\det \u{M}=0$,
\begin{subequations}
  \begin{align}
\det \u{N} &=  \frac{2\tau^*}{4 \Gamma \| \u{D}^*\|}
\left[ (1+B)^2 \xi_1^4 - 2(1-B^2) \xi_1^2 \xi_2^2 + (1-B)^2 \xi_2^4  \right] \,
\label{43a}  \\
 &= \frac{2\tau^*}{4 \Gamma \| \u{D}^*\|} \left[ (1+B) \xi_1^2 - (1-B) \xi_2^2 \right]^2 \, \ge 0 \, ,   \label{43b}
\end{align}
\end{subequations}
and
\begin{equation}
 \chi(\u{M},\u{N}) = \frac{\tau^*}{2  \| \u{D}^*\|} \xi_1^4 + \left( \frac{4}{\Gamma }
+  \frac{\tau^*}{  \| \u{D}^*\|} \right)\xi_1^2 \xi_2^2
+ \frac{\tau^*}{2 \| \u{D}^*\|} \xi_2^4\, >0 \, . \label{42b}
\end{equation}
This proves the lemma.\proofend

\bigskip

{\bf Remark. } It is noteworthy that $\det \u{N}>0$ except for the two directions
\begin{equation}\label{e44}
\frac{\xi_1}{\xi_2} = \pm\sqrt{\frac{1-B}{1+B}}\, .
\end{equation}
Effectively, this calculation rederives the result of Pitman \& Schaeffer \cite{PitmanSchaeffer1987} that the equations of CSSM, even
without $I$-dependence, are well posed for all directions except possibly those defined by \eqref{e44}.

\bigskip

It follows from the lemma that $\u{P}_0 \u{\tilde{U}} = -\Lambda \u{\tilde{U}}$ has two eigenvalues, say
$\Lambda_1,\Lambda_2$, where $\Lambda_1,\Lambda_2<0$ and is homogenous of degree 2 in $\xib$.
Since $\u{P}$ is an $\Oh(|\xib|)$-perturbation of $\u{P}_0$, two of the growth-rate eigenvalues of
\eqref{wellp-second-formula} satisfy
$$
  \lambda_j= \Lambda_j+\Oh(|\xib|)\, , \qquad j=1,2 ,
$$
both of which are negative in the limit $|\xib|\to\infty$; i.e., they are bounded above by zero in this limit.
The third growth rate is given by
$$
  \lambda_3=-\frac{\det \u{P}}{\lambda_1\lambda_2}=  -\frac{\det \u{P}}{\Lambda_1\Lambda_2} +\Oh\left(|\xib|^{-1}\right) .
$$
The first term on the extreme right is the ratio of two quartics, the denominator being nonzero, so it is
bounded, and the perturbation decays at infinity.
This verifies \eqref{wellp-first-formula} for all three eigenvalues derived from \eqref{21a}-\eqref{21e}.

It remains to consider the effect of the lower-order terms that were neglected in \eqref{21a}-\eqref{21e}.
Inclusion of these terms would lead, after a calculation as above, to an eigenvalue
problem \eqref{wellp-second-formula} for a perturbed matrix
$$
  \bthree \fracds{\u{M}+\u{N}}{\rho_*\phi^*} +\Oh(\xib) & & \fracds{i\u{V}}{\rho_* \phi^*} +\Oh(1) \\
    \\
  i \phi^* \xib^T +\Oh(1)& &  \Oh(1)  \ethree .
$$
As above, two of the eigenvalues of this matrix are negative and $\Oh(|\xib|^2)$, and
invoking the determinant shows that the third is bounded. 
This verifies \eqref{wellp-first-formula} for eigenvalues of the full linearization of 
\eqref{12a}-\eqref{12e} and hence shows that the system is linearly well posed.

\section{Conclusions and discussion}

In this paper we have proposed and analysed a synthesis of critical state soil mechanics and the $\mu(I)$-rheology. 
We have found that inclusion of compressibility removes the ill-posedness at low and high inertial numbers in the 
incompressible $\mu(I)$ equations.

Simultaneously, the result shows that the introduction of rate-dependence into CSSM, through variation of the 
inertial number,
gives linearly well-posed equations, provided that the yield locus and flow rule satisfy 
\eqref{Cmu-well-posed-cond} and \eqref{ineq-for-thm}.

\section*{Appendix 1: Ideas from Critical State Soil Mechanics}

\addtocounter{section}{1}

\subsection*{A1.1 Constitutive equations}
Critical State Soil Mechanics (CSSM) is an ingeniously constructed version of plasticity that includes
compressibility but reduces to a singular perturbation of Coulomb material, which is incompressible, in an 
appropriate limit.
In two-dimensional CSSM, flow is described by the usual six variables, $\phi$, $\u{u}$, and $\u{\sigma}$.
Since flow is compressible, the solids fraction $\phi$ remains as a genuine variable.
The governing equations consist of the conservation laws \eqref{c-mass},
\eqref{c-mom} plus three constitutive laws. One of the constitutive equations is the
alignment condition \eqref{align-coulomb}, with no changes required.
The second constitutive equation, like \eqref{yield-cond-coulomb}, specifies the norm of
the deviatoric stress,
\begin{equation} \label{yield-condition-CSSM}
 \mbox{\bf Yield condition:} \qquad \|\u{\tau}\| = Y(p,\phi)  \, ,
\end{equation}
but as indicated the function $Y$ depends on the solids fraction $\phi$ as well as on the mean
stress $p$.
The final constitutive relation, the flow rule, relates expansion and contraction of material to the slope of the
yield surface,
\begin{equation} \label{flow-rule-CSSM}
 \mbox{\bf Flow rule:} \qquad \mbox{div }\u{u}=
2 \frac{\partial Y}{\partial p}(p,\phi) \, \|\u{D}\| \, .
\end{equation}
We refer to Jackson \cite{Jackson1983} for a derivation of \eqref{flow-rule-CSSM} from the normality condition of
plasticity. 

By way of example, a simple, physically acceptable, yield locus is given by
\begin{equation} \label{sample-yield-cond}
 Y(p,\phi) = 2\mu p -p^2/C(\phi) \, ,
\end{equation}
where $\mu$ is a coefficient of friction, as in \eqref{yield-cond-coulomb}, and
$C(\phi)$ is an increasing function of the solids fraction of the form \eqref{form-of-C}.
For such a yield condition,
it follows from the proof of Lemma~3, restricted to the case where $\mu(I)$ is independent of $I$, that
equations \eqref{c-mass}, \eqref{c-mom}, \eqref{yield-condition-CSSM}, \eqref{flow-rule-CSSM}, \eqref{align-coulomb}
reduce to the Coulomb model in the limit $\eps\to0$.

\subsection*{A1.2 Consequences of the flow rule}

The behaviour discussed in this subsection occurs under fairly general hypotheses---see Jackson \cite{Jackson1983}.
However, to explain the theory with a minimum of technicalities, we confine the discussion to the specific
yield condition \eqref{sample-yield-cond}.

The phrase \emph{critical state}, from which CSSM derives its name, refers to a state $p,\phi$ such that
\begin{equation} \label{crit-state-cond}
 \frac{\partial Y}{\partial p}(p,\phi) =0 \, .
\end{equation}
For the example yield condition \eqref{sample-yield-cond}, condition \eqref{crit-state-cond} means that
\begin{equation} \label{crit-state-cond-ex}
  2(\mu-p/C(\phi))=0 \, .
\end{equation}
Rewriting the yield condition as $ Y(p,\tau)=[2\mu-p/C(\phi)]p$ and invoking
\eqref{crit-state-cond-ex}, we deduce that at a critical state
\begin{equation} \label{crit-state-line}
  \|\taub\| = \mu p \, .
\end{equation}
The set where \eqref{crit-state-line} is satisfied is called
the critical state line.
Thus, \emph{along the critical state line, the stress satisfies the Coulomb yield condition.}

According to the flow rule \eqref{flow-rule-CSSM}, at a critical state, deformation is not accompanied by any
change in $\phi$.
Let us examine behaviour away from the critical state line.
Suppose that, for example, initially the (uniform) state of material is at yield at the point $A$ in
Figure~\ref{fig02}(b).
At this point, $\partial Y/\partial p<0$, so according to flow rule $\mbox{div }\u{u}<0$; i.e., material
compactifies and becomes stronger, so $\tau$ must increase for deformation to continue. Indeed, the stress will
continue to increase until a critical state on a larger yield surface is reached, as suggested in the figure by the
$\phi_3$-yield surface. Moreover, if $\eps$ in \eqref{form-of-C} is small, a very slight
increase in $\phi$ is sufficient to accommodate this evolution. I.e., we expect stress to be quickly driven from
the point $A$ to a critical state on a larger yield surface where the Coulomb yield condition
\eqref{crit-state-line} is satisfied.

Conversely, at point $B$ in the figure, $ \partial Y/\partial p>0$, so
under
deformation $\mbox{div }\u{u}>0$; i.e., material expands and becomes weaker.
It is natural to imagine that the stress is driven to a critical
state on a smaller yield surface, as suggested by the arrow in the figure.
This would indeed be the case \emph{if material deformed uniformly},
but this assumption is unrealistic for stresses above the critical state line, $\tau>\mu p$. 
For such stresses, because
material expands under deformation and therefore weakens, instability often causes localised
deformation---if deformation near one point happens to be slightly larger than elsewhere, the associated expansion
lowers the yield condition more near this point, and subsequent deformation tends to concentrate near this point.

\section*{Appendix 2:  A primer on ill posed PDEs}

The following appendix gives a self-contained, elementary summary of key issues regarding ill-posed PDEs.  
A much more detailed treatment can be found in Joseph \& Saut \cite{JosephSaut1990}.

\addtocounter{section}{1}
\subsection*{A2.1 Testing for ill-posedness}

The initial value problem for a PDE is called \emph{well posed} in the sense of Hadamard if for general initial data a solution (1)~exists, (2)~is unique and (3)~varies continuously under perturbations of the initial conditions\footnote{More precisely regarding Condition~(1): we choose a positive integer $k$ and require that the 
IVP has a solution for any initial conditions in $\mathcal{BC}^k(\R^n)$, i.e., for
$k$-times continuously differentiable functions such that all 
derivatives of order $k$ or less are bounded. 
Likewise regarding Condition~(3): we require that for the same integer $k$ and 
for any positive time $T$, the solution operator is continuous as a map from $\mathcal{BC}^k(\R^n)$
into continuous functions on  $[0,T]\times\R^n$.
We refer to Joseph \& Saut \cite{JosephSaut1990} for elaboration of these issues.}  (cf. also Pinchover \& Rubinstein \cite{PinchoverRubinstein2005}). 
If one or more of these criteria is not satisfied then the problem is called \emph{ill posed}. A classic example of an 
ill-posed problem is the backward heat equation
\begin{equation} \label{backward-heat}
  \partial_t u=-\partial_{xx}u \, .
\end{equation}
In Section~A2.2 below we show Condition~(1) fails; here we show Condition~(3) also fails.
Taking the Fourier transform reveals that the equation admits solutions
\begin{equation} \label{backward-heat-growing-sln}
 u_\xi(x,t) = \sin(\xi x)e^{\xi^2 t}\, ,
\end{equation}
for any $ \xi \in \R$. 
Consider the scaled solutions $|\xi|^{-p}  u_\xi(x,t)$, where $p>0$, as perturbations of the trivial solution $u(x,t)\equiv0$.
The initial conditions of the scaled solution---i.e., $|\xi|^{-p} \sin(\xi x)$---tend to zero in the 
sup norm as $\xi\to\infty$; 
indeed, if $p>k$ these initial conditions tend to zero in the $\mathcal{C}^k$ norm.
On the other hand, for any $t>0$ the norm
\begin{equation} \label{backward-heat-blow-up}
  \sup_{x\in\R} |\xi|^{-p} u_\xi(x,t) 
\end{equation}
tends to infinity in this limit.
Thus, an arbitrarily small perturbation of initial conditions for \eqref{backward-heat} can lead to an 
arbitrarily large solution in an arbitrarily short time.

For more general PDEs there is a test for ill-posedness based on Fourier analysis of the 
\emph{linearisation} of the equations. The process is summarised as:
\begin{enumerate}
\item Linearise the equations about a base-state solution;
\item Freeze the coefficients at some point $(\u{x}^*,t^*)$;
\item Look for solutions with exponential dependence 
  $e^{i\langle \u{\xi},\u{x}\rangle +\lambda(\u{\xi})t} \,.$ 
\end{enumerate}
We shall say the original PDE is linearly ill-posed (with respect to the base-state solution 
at the given point) if
$$
  \limsup_{|\xib|\to\infty} \lambda(\xib) = +\infty .
$$
For most examples, if a PDE is linearly ill-posed, it is 
ill posed in the sense of Hadamard. (But see Kreiss \cite{Kreiss1978} for exceptional examples.)

An equation is called \emph{linearly well-posed} with respect to a given base solution if the growth rate 
is bounded from above for all points $(\u{x}^*,t^*)$. 
Linear well-posedness does not imply well-posedness in the sense of Hadamard.
For example, it is trivially verified that the Navier-Stokes equations are linearly well-posed, 
but a major effort is required to show that, even just for a finite time, they are well posed in the sense of Hadamard, 
and it is not known whether they are well posed for all time.

We illustrate the above test on the following made-up nonlinear system that has some similarity to the PDEs 
analysed in this paper, 
\begin{equation}\label{toy-pde}
\begin{array}{ccl}
\d{t}u=\d{x}v \, , \\
\d{t}v=\varepsilon u \d{{xx}}v + \d{x}u - v - \sin(x) \, .
\end{array}
\end{equation}
The linearised equations with frozen coefficients are 
\begin{equation}\label{toy-pde-full-lin}
\begin{array}{ccl}
\d{t}\hat{u}=\d{x}\hat{v} \, , \\
\d{t}\hat{v}=\varepsilon \left[ u^* \;\d{{xx}}\hat{v}+ (\d{{xx}}v)^*\, \hat{u}\right] + \d{x}\hat{u} -\hat{v}  \, ,
\end{array}
\end{equation}
where $u^*, v^*$ is the base-state solution evaluated at the point $(x^*,t^*)$ and $\hat{u},\hat{v}$ are 
the perturbations. 
These constant-coefficient linear PDEs have exponential solutions
\begin{equation} \label{toy-pert}
\u{\hat{U}}=\bvec \hat{u} \\ \hat{v} \evec= e^{i\xi x+\lambda t} \u{\tilde{U}} \, ,
\end{equation} 
where $\u{\tilde{U}}\in\R^2$ satisfies the eigenvalue problem
\begin{equation} \label{toy-mat}
\btwo 0 & i\xi \\ i\xi+\eps \d{{xx}} v^* & -\varepsilon u^* \xi^2 -1 \etwo\u{\tilde{U}}=\lambda \u{\tilde{U}} \, .
\end{equation} 
The eigenvalues of \eqref{toy-mat} could be easily calculated exactly but, provided that $u^*\neq 0$, they
can be estimated more easily from their 
asymptotic behaviour as $|\xi|\to\infty$:
\begin{equation} \label{toy-eig}
\lambda_1 = - \varepsilon u^* \xi^2 + \Oh(\xi)
\quad \mbox{and} \quad \lambda_2= \frac{\det S(\xi)}{\lambda_1}=
-\frac{1}{\varepsilon u^*} + \Oh(\xi^{-1}) \, .
\end{equation}
where $S(\xib)$ is the $2\times2$ matrix in \eqref{toy-mat}.
If $u^*>0$, then the eigenvalues satisfy 
\begin{equation} \label{cond-well-p}
 \max_{j=1,2} \; \limsup_{|\xi|\to\infty} \; \Re \lambda_j(\xi) < \infty \, ,
\end{equation}
so \eqref{toy-pde} is linearly well posed.
On the other hand if
$u^*<0$, then $\lambda_1$ is unbounded, so \eqref{toy-pde} is ill posed.

Note that in analysing linear ill-posedness of \eqref{toy-pde} we consider the 
\emph{full} linearization of the equations, i.e., \eqref{toy-pde-full-lin}. 
One might be tempted to discard terms with lower-order derivatives in the expectation that the 
growth of exponential solutions as $|\xi|\to\infty$ ought to be dominated by the highest-order
derivatives in the equation.
However, the counter-example
$$
  \partial_t u = \partial_{xxx} u - \partial_{xx} u
$$
shows this expectation is not valid in general.

Nevertheless, for this example we may in fact analyse exponential solutions of \eqref{toy-pde-full-lin} by first considering the 
\emph{maximal order} linearised equations
\begin{equation}\label{toy-pde-lin}
\begin{array}{ccl}
\d{t}\hat{u}=\d{x}\hat{v} \, , \\
\d{t}\hat{v}=\varepsilon u^* \d{{xx}}\hat{v} + \d{x}\hat{u}  \, .
\end{array}
\end{equation}
In each equation of \eqref{toy-pde-lin}, only terms of maximal order are retained, i.e.,
the terms $(\d{{xx}} v)^* \hat{u}$ and $\hat{v}$ have been dropped from the second equation
because it contains
the higher-order terms $\d{x}\hat{u}$ and $u^* \;\d{{xx}}\hat{v}$, respectively.
The growth rate of exponential solutions of \eqref{toy-pde-lin} satisfy the same estimates \eqref{toy-eig},
and the neglected lower-order terms don't change the leading-order behaviour.
Often calculations may be simplified by studying the maximal-order linearization as an intermediate step.

\subsection*{A2.2 Consequences of ill-posedness}

\subsubsection*{A2.2.1 Restrictions on the existence of solutions}

In order for an ill-posed initial value problem to have a solution, usually the initial conditions must satisfy an
extreme smoothness requirement, stronger than is physically acceptable in most applications.
It may be difficult to demonstrate this behaviour in general, but for the backwards heat equation, we illustrate
the behaviour with

\bigskip
\noindent{\bf Proposition.} \emph{If $\eps\neq0$, the initial value problem for \eqref{backward-heat} with
initial data $u(x,0)=\eps|\sin x|^p$ has no solution for any positive time interval unless the power $p$ is an
even non-negative integer.%
\footnote{Of course the general solution of \eqref{backward-heat} is a linear superposition of the 
solutions \eqref{backward-heat-growing-sln}.  We have no need for the general solution since one 
counterexample is sufficient to invalidate Condition~(1) above.}}

\bigskip

\noindent{\bf Proof.}
Suppose  the $2\pi$-periodic function $f(x)$ has Fourier series
$
  f(x) \sim \sum \; c_n e^{inx} .
$
If equation \eqref{backward-heat} with initial condition $u(x,0)=f(x)$ has a continuous solution for $0\le
t<\eta$, it has the Fourier-series representation
$$
u(x,t) = \sum_{n=-\infty}^\infty \; c_n e^{inx+n^2 t} \;.
$$
Moreover for $0\le t<\eta$,
\begin{equation} \label{L2norm-sln}
 \sum_{n=-\infty}^\infty \; e^{2n^2 t} |c_n|^2=
\frac{1}{2\pi} \int_{-\pi}^\pi |u(x,t)|^2 \, dx <\infty\, .
\end{equation}

Now if  the Fourier coefficients of a function $f(x)$ satisfy
$
  \sum \; n^{2k} |c_n|^2   <\infty \, ,
$
then the derivatives  $(d/dx)^j f(x)$ are square integrable for $j=0,1,\ldots,k$.
But, provided $p$ is not an even non-negative integer, the proposed initial data $|\sin x|^p$ has singular
behaviour
near $x=0$.
Specifically, $(d/dx)^k |\sin x|^p$ is square integrable only if
$k<p+\half$.
It follows for the Fourier coefficients of $|\sin x|^p$ that if $k>p+\half$,
$$
  \sum_{n=-\infty}^\infty \; n^{2k} |c_n|^2   = \infty \,.
$$
This inequality is incompatible with \eqref{L2norm-sln}, so the initial value problem cannot be solved on any
positive time interval.\proofend

\subsubsection*{A2.2.2 Grid-dependent computations}

The attempt to solve an ill posed PDE numerically produces unreliable, grid dependent, results.
Such behaviour has been observed in various physical problems \cite{Gray1999, Woodhousetal2012, 
Barker2015} where the formulation was based on an ill-posed system of equations. However, in 
complicated problems like these, usually computational resources are stretched to the limit, meaning behaviour 
under grid refinement cannot be readily probed. Let us illustrate grid dependence on a much less 
demanding problem, the toy problem \eqref{toy-pde} above. 

\begin{figure}
\begin{center}

\SetLabels 
      \L (0.0*0.9) $(a)$\\
      \L (0.5*0.9) $(b)$\\
      \E (0.01*0.5) \rotatebox{90}{$\mbox{Dist}$}\\
      \L (0.25*0.0) $t$\\
            \E (0.51*0.5) \rotatebox{90}{$u,v$}\\
      \L (0.75*0.0) $x$\\
       \endSetLabels
       \strut\AffixLabels{
	\includegraphics[width=0.5\textwidth]{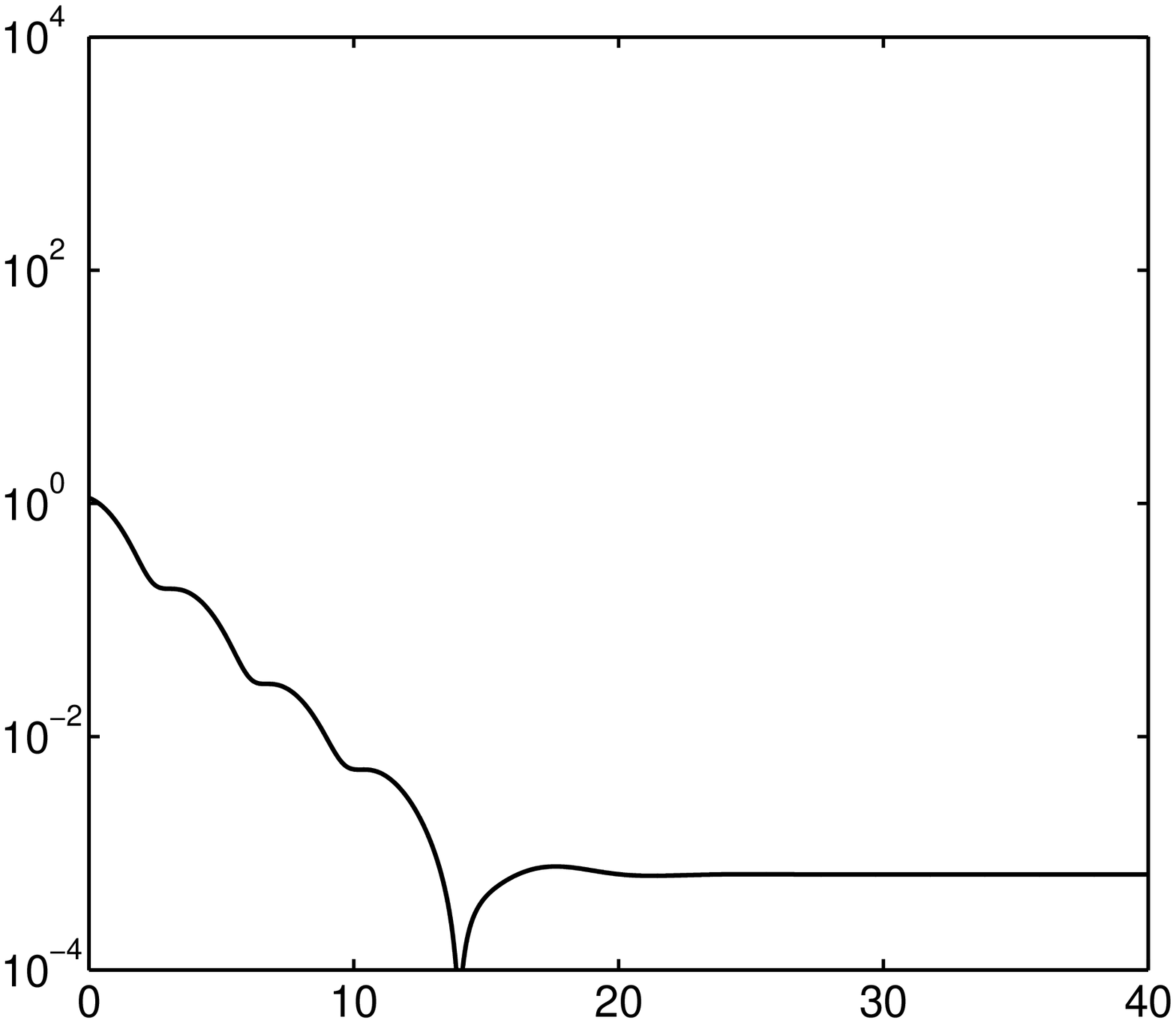} 
	\includegraphics[width=0.5\textwidth]{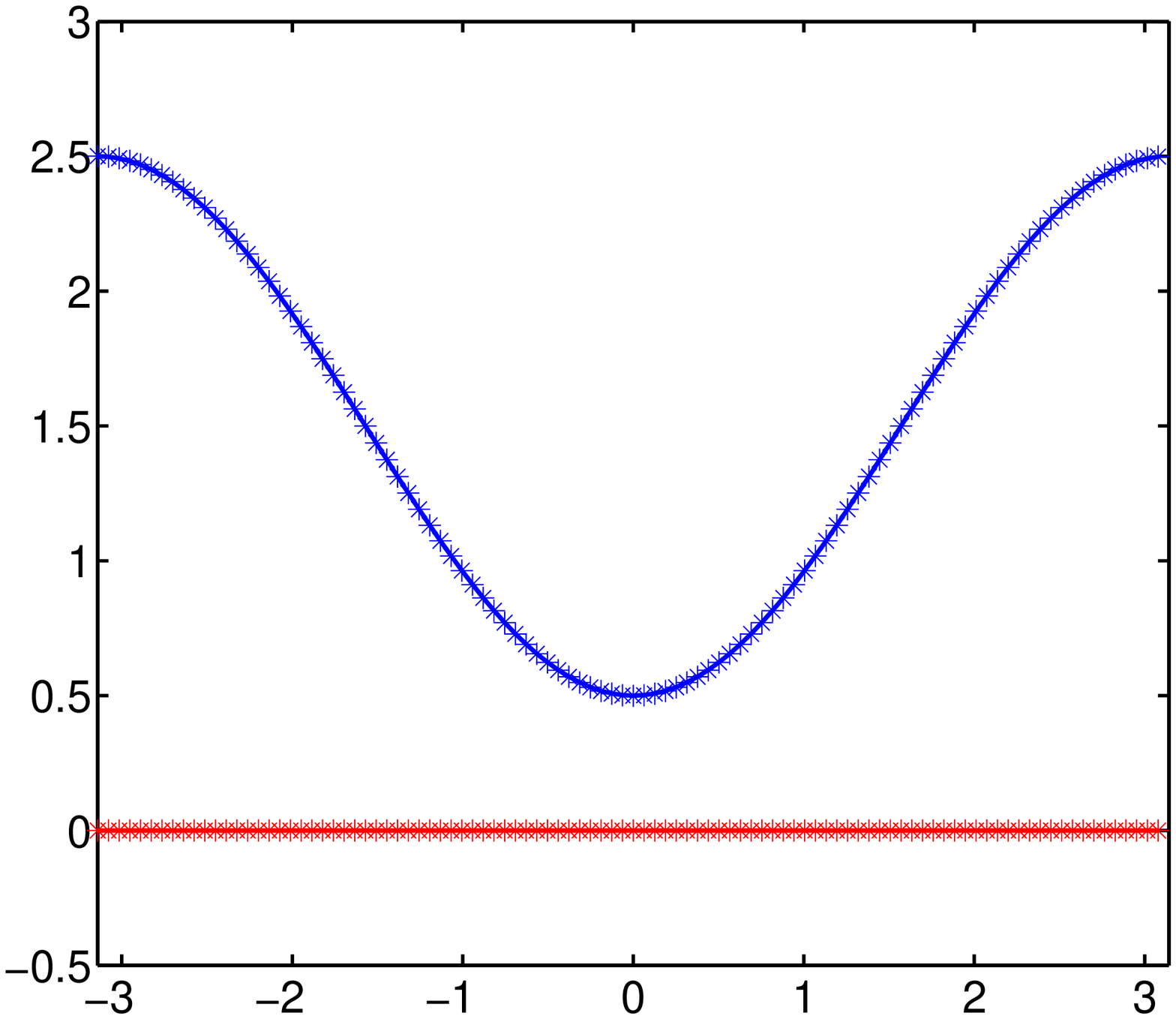}
	}

\caption{  \label{fig31} Numerical solutions of \eqref{toy-pde} with the distance from the asymptotic solution \eqref{toy-err} in $(a)$ and the fields at $t=100$ in $(b)$. Here $a=1.5$, $\varepsilon=0.01$ and the discretisation is $\Delta x = 2\pi / 100$ and $\Delta t = 1 \times 10 ^{-3}$.}
\end{center}
\begin{center}
\SetLabels
      \L (0.0*0.9) $(a)$\\
      \L (0.5*0.9) $(b)$\\
      \E (0.01*0.5) \rotatebox{90}{$\mbox{Dist}$}\\
      \L (0.25*0.0) $t$\\
            \E (0.51*0.5) \rotatebox{90}{$u,v$}\\
      \L (0.75*0.0) $x$\\
       \endSetLabels
       \strut\AffixLabels{
	\includegraphics[width=0.5\textwidth]{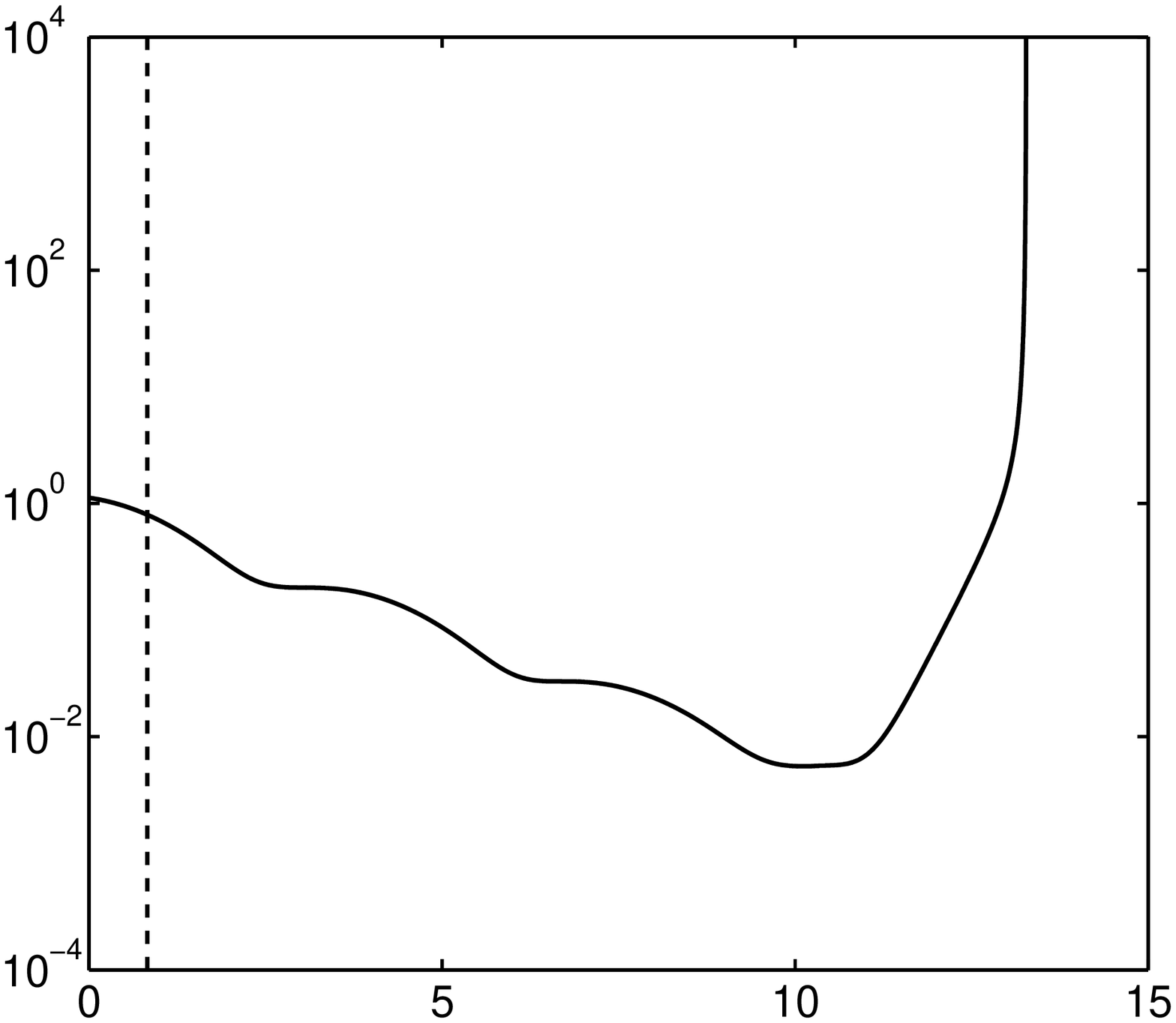} 
	\includegraphics[width=0.5\textwidth]{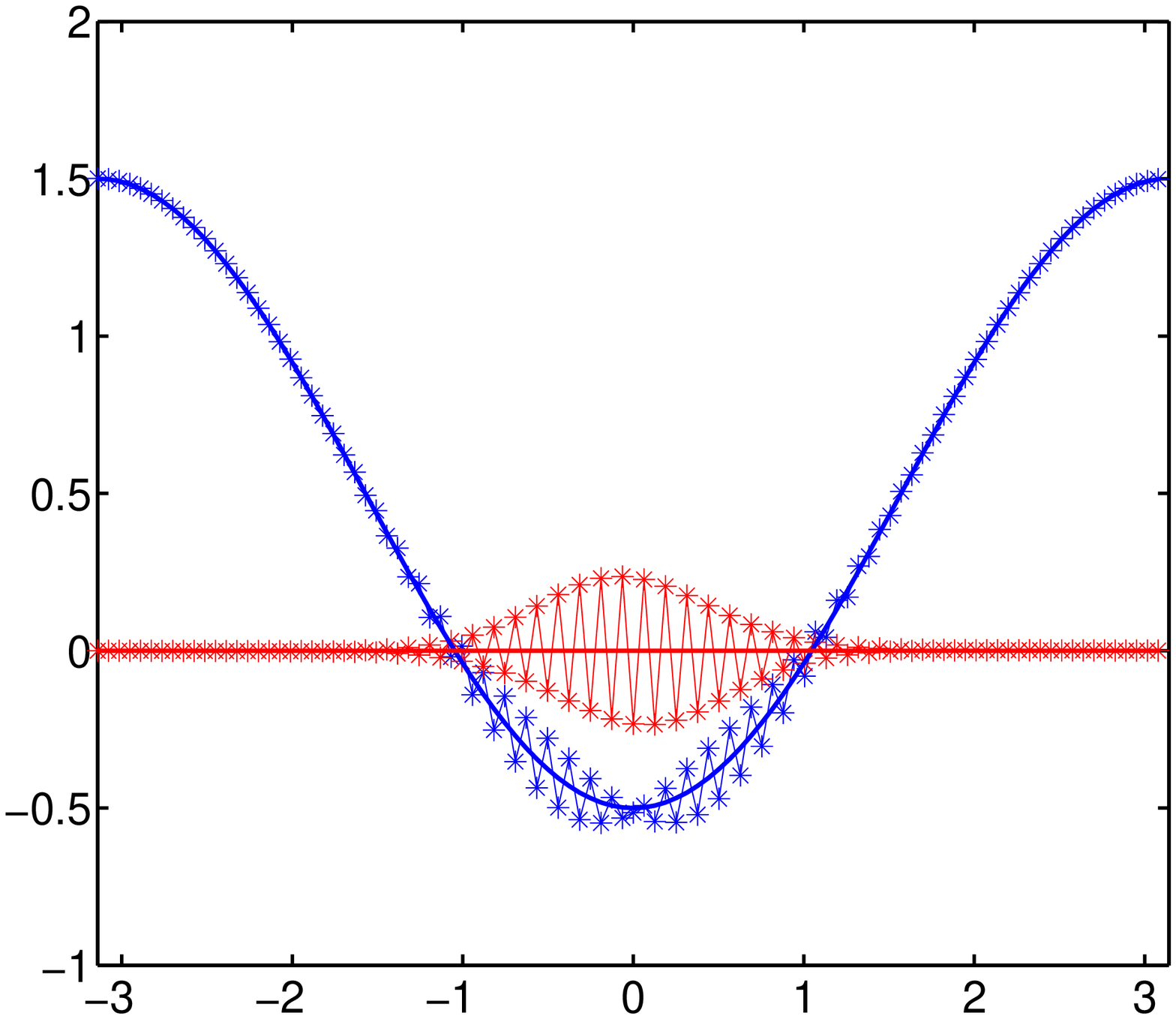}
	}
\caption{t  \label{fig33}Numerical solutions of \eqref{toy-pde} with the distance from the asymptotic solution 
\eqref{toy-err} 
in $(a)$ and the fields at $t=12.5$ in $(b)$. Here $a=0.5$, $\varepsilon=0.01$ and the discretisation is $\Delta x 
= 2\pi / 100$ and $\Delta t = 1 \times 10 ^{-3}$. The vertical dashed line in $(a)$ is the first time that $u=0$.}
\end{center}
\end{figure}

If $\eps=0$ and with initial conditions
\begin{equation} \label{toy-initial}
u(x,0) = a \, , \qquad  v(x,0)=-\sin(x)/2 \, ,
\end{equation}
the (linear) equations  \eqref{toy-pde} have the exact solution 
\begin{equation} \label{toy-sln}
\begin{array}{ccl}
 u(x,t)&	=&	\frac{e^{-t/2}}{2} [\cos(x+\sqrt{3}t/2)+\cos(x-\sqrt{3}t/2)] +a -\cos x\\
 v(x,t)&	=&	\frac{e^{-t/2}}{2} [\cos(x+\sqrt{3}t/2+\pi/6) -\cos(x-\sqrt{3}t/2-\pi/6)] \, .
\end{array}
\end{equation}
The large-time limit of these solutions, 
\begin{equation} \label{toy-steady}
u(x,\infty)=a-\cos x \, , \qquad  v(x,\infty)=0\, ,
\end{equation}
is also a steady-state solution of the nonlinear system (with $\eps>0$).
If $a>1$, then $u(x,\infty)>0$, and the calculations above suggest that the equations will be linearly 
well-posed. 
However if $a<1$, then $u(x,\infty)$ dips below zero over an interval, 
which suggests that the equations will be ill-posed. 

Figure~\ref{fig31}, where $a=1.5$, and Figure~\ref{fig33}, where $a=0.5$, confirm these expectations.
They show numerical solutions using a central-space forward-time explicit scheme on the 
periodic domain $x\in[-\pi,\pi]$ with a spatial resolution of $\Delta x = 2\pi/100$.  
Each figure has two panels, one showing the temporal evolution of the 
distance from the asymptotic solution,
\begin{equation}\label{toy-err}
\mbox{Dist} = \max\left\{\sqrt{|u-u(x,\infty)|^2+|v-v(x,\infty)|^2}\right\} \, ,
\end{equation}
and the other plotting the two variables $u, v$ at a specific (late) time during the computation. 
In Figure \ref{fig31}, the well posed case, the
numerical solution converges to the predicted steady state solution \eqref{toy-steady} within numerical 
accuracy.
By contrast, in Figure~\ref{fig33}, after an initial decay, ill-posedness asserts itself 
and causes the solution to blow up.

\begin{figure}
\begin{center}
\SetLabels 
      \L (0.0*0.9) $(a)$\\
      \L (0.5*0.9) $(b)$\\
      \E (0.01*0.5) \rotatebox{90}{$\mbox{Dist}$}\\
      \L (0.25*0.0) $t$\\
            \E (0.51*0.5) \rotatebox{90}{$u,v$}\\
      \L (0.75*0.0) $x$\\
       \endSetLabels
       \strut\AffixLabels{
	\includegraphics[width=0.5\textwidth]{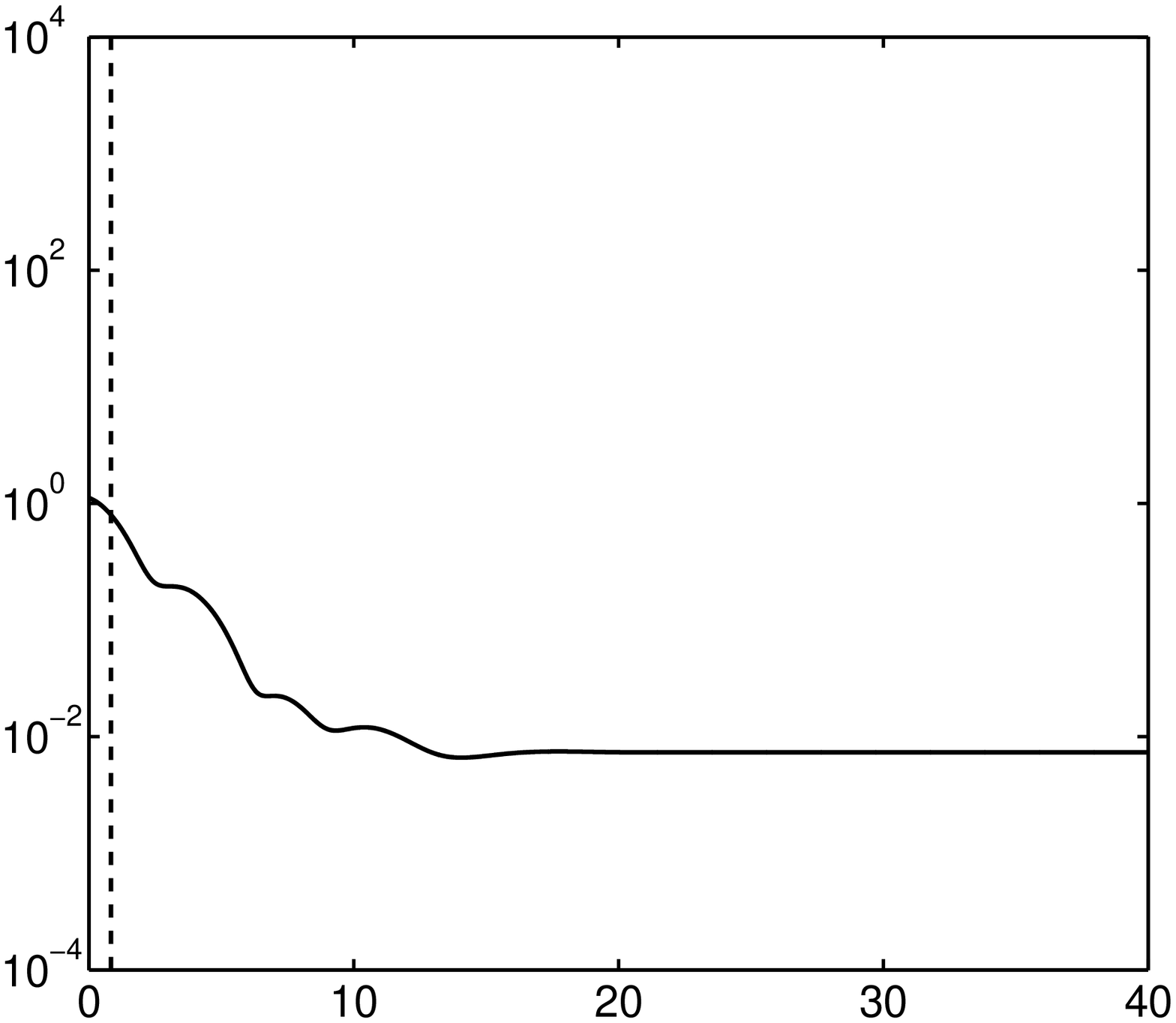} 
	\includegraphics[width=0.5\textwidth]{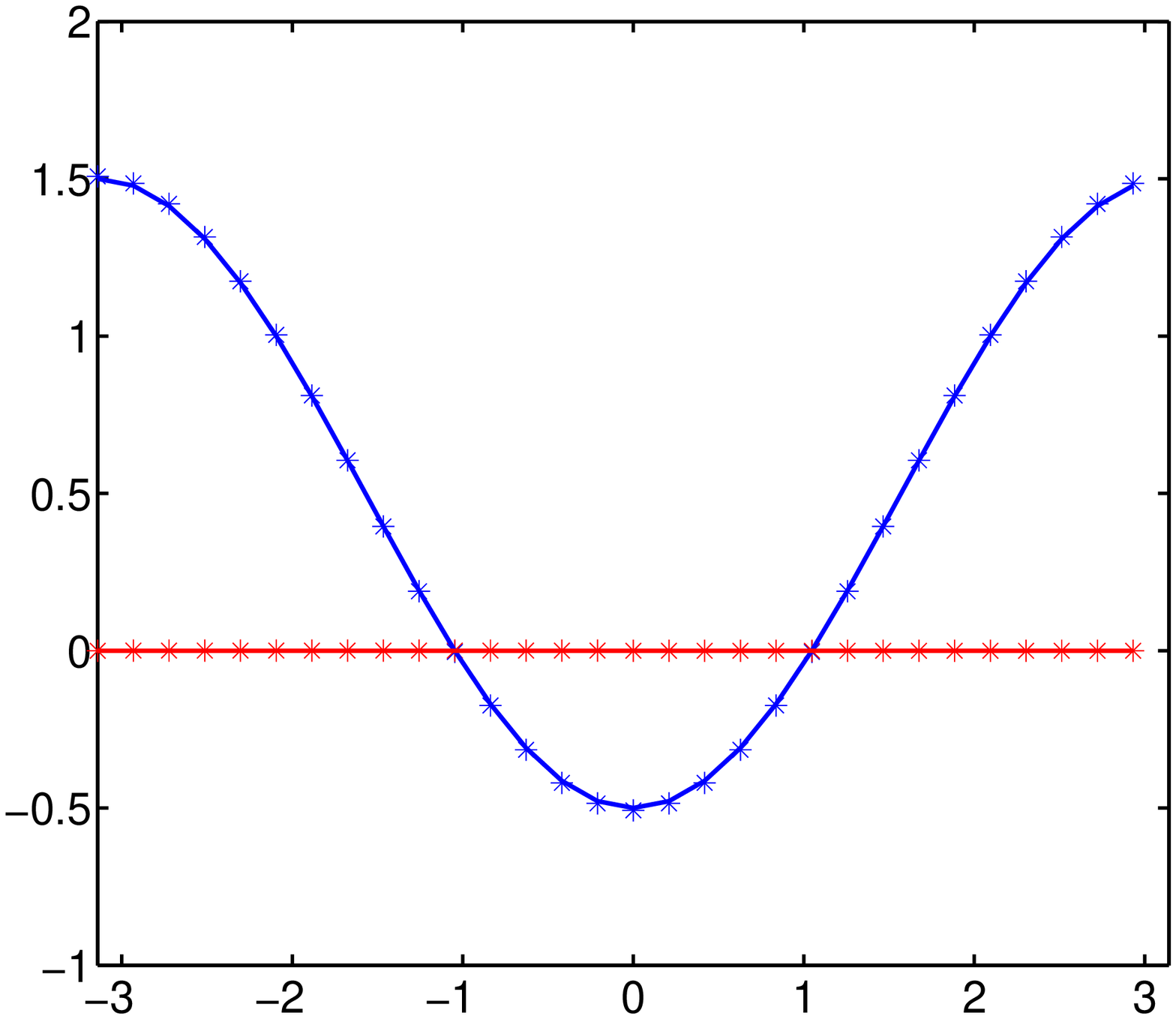}
	}
\caption{  \label{fig32} Numerical solutions of \eqref{toy-pde} with the distance from the asymptotic solution \eqref{toy-err} in $(a)$ and the fields at $t=100$ in $(b)$. Here $a=0.5$, $\varepsilon=0.01$ and the discretisation is $\Delta x = 2\pi / 30$ and $\Delta t = 1 \times 10 ^{-3}$. The vertical dashed line is the first time that $u=0$.}
\end{center}
\end{figure}

Regarding grid dependence,
Figure~\ref{fig32} shows another computation in the ill posed case with a coarser grid, $\Delta x = 2\pi/30$.
The solution appears to converge to the steady state solution, just like in the well posed case.
In other words, the computations on the coarse grid hide the ill posed character of the underlying PDEs. This 
highlights that in order to extract meaningful information from numerical computations, a proper study of grid 
convergence must be first carried out. 

Incidentally, if the grid is made finer than in Figures~\ref{fig31} and \ref{fig33}, in the well posed case
$a>1$ the numerical solution converges to the steady-state solution with a smaller numerical error,
while in the ill-posed case it blows up \emph{sooner}, as expected.

\par
\qquad
\par

\subsection*{Acknowledgements}
This research was supported by NERC grants NE/E003206/1 and NE/K003011/1 as
well as EPSRC grants EP/I019189/1, EP/K00428X/1 and EP/M022447/1.  J.M.N.T.G. is
a Royal Society Wolfson Research Merit Award holder (WM150058) and an EPSRC
Established Career Fellow (EP/M022447/1). Research of M.S. was supported by National Science Foundation grant DMS-1517291.

\bibliography{critical}

\end{document}